\documentclass[twocolumn]{aastex631}

\pdfoutput=1 
\usepackage{natbib}
\let\tablenum\relax
\usepackage{siunitx}
\usepackage{gensymb}
\usepackage{amssymb}
\usepackage{graphicx}
\usepackage[super]{nth}
\usepackage[version=4]{mhchem}
\begin{document}

\title[Coy et al.: Trend in Brightness Temperatures of M-Earths]{Population-level Hypothesis Testing with Rocky Planet Emission Data: A Tentative Trend in the Brightness Temperatures of M-Earths}

%Alternate title if not focusing on trend: ???

\author[0000-0002-0508-857X]{Brandon Park Coy}
\affiliation{Department of the Geophysical Sciences, University of Chicago, Chicago, IL, USA}
\correspondingauthor{Brandon Park Coy}
\email{bpcoy@uchicago.edu}

\author[0000-0003-2775-653X]{Jegug Ih}
\affiliation{Space Telescope Science Institute, Baltimore, MD, USA}
\affiliation{Department of Astronomy, University of Maryland, College Park, MD, USA}

\author[0000-0002-1426-1186]{Edwin S. Kite}
\affiliation{Department of the Geophysical Sciences, University of Chicago, Chicago, IL, USA}

\author[0000-0002-9076-6901]{Daniel D.B. Koll}
\affiliation{Department of Atmospheric and Oceanic Sciences, Peking University, Beijing, People's Republic of China}

\author[0000-0003-2802-9561]{Moritz Tenthoff}
\affiliation{Department of Electrical Engineering and Information Technology, Technische Universität Dortmund, Dortmund, Germany}

\author[0000-0003-4733-6532]{Jacob L. Bean}
\affiliation{Department of Astronomy \& Astrophysics, University of Chicago, Chicago, IL, USA}

\author[0000-0003-4241-7413]{Megan Weiner Mansfield}
\affiliation{School of Earth and Space Exploration, Arizona State University, Tempe, AZ, USA}
\affiliation{Department of Astronomy, University of Maryland, College Park, MD, USA}

\author[0000-0002-0659-1783]{Michael Zhang}
\affiliation{Department of Astronomy \& Astrophysics, University of Chicago, Chicago, IL, USA}

\author[0000-0002-6215-5425]{Qiao Xue}
\affiliation{Department of Astronomy \& Astrophysics, University of Chicago, Chicago, IL, USA}

\author[0000-0002-1337-9051]{Eliza M.-R. Kempton}
\affiliation{Department of Astronomy, University of Maryland, College Park, MD, USA}

\author[0000-0002-4324-1459]{Kay Wohlfarth}
\affiliation{Department of Electrical Engineering and Information Technology, Technische Universität Dortmund, Dortmund, Germany}

\author[0000-0003-2215-8485]{Renyu Hu}
\affiliation{Jet Propulsion Laboratory, Pasadena, CA, USA}
\affiliation{Division of Geological and Planetary Sciences, California Institute of Technology, Pasadena, CA, USA}

\author[0009-0004-9766-036X]{Xintong Lyu}
\affiliation{Department of Atmospheric and Oceanic Sciences, Peking University, Beijing, People's Republic of China}

\author{Christian Wöhler}
\affiliation{Department of Electrical Engineering and Information Technology, Technische Universität Dortmund, Dortmund, Germany}

\begin{abstract}
\noindent Determining which rocky exoplanets have atmospheres, and why, is a key goal for the James Webb Space Telescope. So far, emission observations of individual rocky exoplanets orbiting M stars (M-Earths) have not provided definitive evidence for atmospheres. Here, we synthesize emission data for M-Earths and find a trend in measured brightness temperatures (ratioed to its theoretical maximum value) as a function of instellation.  However, the statistical evidence of this trend is dependent on the choice of stellar model, and we consider its identification tentative. We show that this trend can be explained by either the onset of thin/tenuous ($<1$ bar) CO$_2$-rich atmospheres on colder worlds, or a population of bare rocks with stronger space weathering and/or coarser regolith on closer-in worlds. Such grain coarsening may be caused by sintering near the melting point of rock or frequent volcanic resurfacing. Furthermore, we highlight considerations when testing rocky planet hypotheses at the population level, including the choice of instrument, stellar modeling, and how brightness temperatures are derived.
We also find that fresh (unweathered) fine-grained surfaces can serve as a false positive to the detection of moderate atmospheric heat redistribution through eclipse observations.   However, we argue that such surfaces are unlikely given the ubiquity of space weathering in the solar system, the low albedo of solar system airless bodies, and the high stellar wind environments of M-Earths. 
Emission data from a larger sample of M-Earths will be able to confirm or reject this tentative trend and diagnose its cause through spectral characterization. 
\end{abstract}

\keywords{Exoplanets (498), James Webb Space Telescope (2291), Exoplanet atmospheres (487), Extrasolar rocky planets (511)}

\section{Introduction} \label{sec:intro}

A key goal for the James Webb Space Telescope (JWST) is to determine the prevalence and origins of terrestrial (i.e., ``rocky'', smaller than $\SI{1.5}{R_{\oplus}}$) exoplanet atmospheres.  Toward this goal, studying terrestrial exoplanets orbiting M dwarfs (``M-Earths") is crucial, as these stars are the most abundant in our neighborhood and their favorable signal sizes are currently the most amenable to investigating potentially habitable worlds \citep{seager13, barstow16,lustig19}.  

Observing the thermal emission of tidally-locked rocky exoplanets in secondary eclipse allows for efficient detection of atmospheres on such worlds \citep{koll19,mansfield19}.  Secondary eclipse observations allow for measuring a planet's dayside temperature, which can constrain the amount of incident heat redistributed to the nightside by the atmosphere.  As a thick atmosphere should cool the dayside by redistributing heat over the entire planet, observing a colder dayside than expected for a `bare-rock' planet can indicate the presence of an atmosphere. This effect is evident in the nontidally locked solar system terrestrial planets (e.g., see Fig. 4 of \citealt{xue24}). Additionally, high-albedo cloud decks, like those on Venus, would further cool the planet's dayside.  

Beyond their detection, characterizing the atmosphere is also potentially viable.  As many gas species expected in rocky planet atmospheres (e.g., \ce{CO2}, \ce{H2O}, \ce{SO2}) are infrared absorbers, they both control the vertical thermal structure of the atmosphere and create spectral features in thermal emission.  As such, spectroscopic or multi-band emission observations can simultaneously constrain the thickness of the atmosphere and its composition (e.g., \citealt{deming09, whittaker22}).

On the other hand, if the planet is airless, thermal emission probes surface properties, such as its mineral composition, level of roughness across multiple scales, and degree of space weathering \citep{hu12, whittaker22, lyu2024,tenthoff24,first2024,paragas25}. Thus, JWST thermal emission observations provide the very first pathway towards probing exoplanet surfaces at a population level.

Under what conditions M-Earths could have atmospheres is unclear.  In this context, the search for atmospheres on M-Earths can be framed as constraining the conditions in which such atmospheres can exist, and this framing has been invoked in, e.g., the Rocky Worlds Director's Discretionary Time (DDT) program\footnote{\url{https://rockyworlds.stsci.edu/}} \citep{Redfield2024}.  Whether solar system empirical trends in atmosphere presence/absence can be extrapolated to M stars (i.e., the `Cosmic Shoreline' hypothesis, \citealt{zahnle17}) is currently unknown, and theory suggests the harsh stellar environment of M-Earths is likely inhospitable for atmospheres (e.g., \citealt{davenport12,shields16,dong18} --- we will discuss this point in more detail in Section \ref{sec:discussion}). However, these simple scaling laws do not account for the possible diversity in initial volatile content of rocky worlds, which remains poorly understood.  For example, there is no consensus explanation for Earth's carbon, nitrogen, or water budget (e.g., \citealt{kiteschaefer21,li2021,hirschmann2021,krijt2022}), and volatile loss on rocky exoplanets is also not well-understood (e.g., \citealt{kite20,nakayama22}).

Observations so far using secondary eclipses paint a murky picture for the prospect of discovering atmospheres on M-Earths.  To date, no M-Earths observed in thermal emission have been revealed to conclusively have a thick atmosphere.  All measured dayside temperatures are $1\,\sigma$ consistent with the no-atmosphere (`bare-rock') scenario and $2\,\sigma$ consistent with a zero-albedo blackbody across a wide range of irradiation \citep{kreidberg19, Crossfield22, greene23, zieba23, zhang24, xue24, mansfield24,luque24,Wachiraphan24}, or are plagued by systematics on the order of the expected eclipse depth, complicating interpretation \citep{august2024}.  Moreover, molecular features have not been detected in spectral emission observations for M-Earths \citep{zhang24,xue24,mansfield24,luque24,Wachiraphan24}. 
However, atmospheres remain possible in some cases; e.g., TRAPPIST-1 c may have a 100 mbar Earth-like atmosphere \citep{zieba23,lincowski2023} and TRAPPIST-1 b shows different brightness temperatures at 12.8 and 15 \si{\micron}, possibly indicative of a thermally-inverted \ce{CO2}-rich atmosphere \citep{ducrot23}.  

An alternative approach to interpreting secondary eclipse depths is by examining these observations at the population level to test hypotheses that explain possible trends in the global population. For example, the Cosmic Shoreline hypothesis predicts that more massive and less irradiated exoplanets have thicker atmospheres \citep{zahnle17}. Thus, homogeneously studying population-level emission data can aid in understanding controls on atmosphere presence on M-Earths.

Here, we present a tentative 1D trend in the brightness temperatures of M-Earths as a function of their irradiation temperatures.  We summarize the currently available emission observations in Section \ref{sec:methods}. We present and statistically evaluate the trend in Section \ref{sec:results},  examining a variety of atmospheric and geologic hypotheses that could explain this trend. In Section \ref{ap:additional}, we explore various additional geologic and atmospheric processes that may affect brightness temperatures but are unlikely to explain the proposed trend.  We discuss these planets and future JWST targets in the context of the Cosmic Shoreline hypothesis in Section \ref{sec:discussion}.

\section{Methods}\label{sec:methods}

\subsection{Emission Observations of Rocky Planets}

M-Earths observed in emission thus far span a wide range of irradiation temperatures, which we define as
\begin{equation}
T_{irr} = T_{\star} \sqrt{\frac{R_{\star}}{a}},
\end{equation}  
\noindent where $T_{\star}$ is the host star effective temperature, $R_{\star}$ is the host star radius, and $a$ is the planet's orbital semi-major axis.  The irradiation temperature is $\sqrt{2}$ times the equilibrium temperature of a zero-albedo planet with global heat redistribution, and it is equivalent to the substellar temperature of a tidally-locked world with zero albedo at all wavelengths (i.e., a perfect blackbody).  Irradiation temperatures of planets considered here range from the molten sub-Earth GJ 367 b ($T_{irr}=\SI{1930}{K}$) to the potential ``Venus twin'' TRAPPIST-1 c ($T_{irr}=\SI{480}{K})$.

The expected disk-integrated dayside temperature of a planet can be calculated from the planet's Bond albedo $A_{B}$ and a heat redistribution factor $f$ \citep{hansen08,cowan11},

\begin{equation}
    T_{d}=T_{irr} f^{1/4}(1-A_{B})^{1/4},
\end{equation}
where $f \rightarrow \frac{2}{3}$ for a planet with zero heat redistribution to the nightside to $f \rightarrow \frac{1}{4}$ for full heat redistribution.  $f$ is also often commonly rewritten as $\varepsilon$, defined as
\begin{equation}
    f=\frac{2}{3}-\frac{5}{12}\varepsilon,
\end{equation}
so that $\varepsilon$ varies from 0 with no heat redistribution to 1 with full redistribution.  Thus, the theoretical maximum disk-integrated dayside temperature for a zero-albedo, zero-heat redistribution planet is

\begin{equation}
    T_{d,max}=T_{irr}\left(\frac{2}{3}\right)^{1/4}.
\end{equation}

Following recent works \citep{xue24,mansfield24,Wachiraphan24}, we define the `brightness temperature ratio' $\mathcal{R}$ as:

\begin{equation}
    \mathcal{R}\equiv \frac{T_{d}}{T_{d,max}},
\end{equation}
which compares the measured dayside brightness temperature to the theoretical maximum.   Another useful metric is the inferred (or `effective') albedo, which is the planetary Bond albedo needed to reproduce the observed dayside temperature assuming zero heat redistribution and unit emissivity and can be calculated via,
\begin{equation}
    A_{i}=1-\mathcal{R}^{4}.
\end{equation}

\subsection{Determining Brightness Temperature Ratios}

\begin{deluxetable*}{lcccccccc}
\tablecaption{Host Star and System Parameters Used in Brightness Temperature Nested Sampling}
\tablewidth{0pt}
\tablehead{
\colhead{Star} & \colhead{$T_{\star}$ (K)} & \colhead{log($g[cm/s^{2}]$)} & \colhead{[M/H]} & \colhead{$a/R_{\star}$} & 
\colhead{$R_{p}/R_{\star}$}}
\startdata
    TRAPPIST-1 c & $2566\pm26$ & $5.2395^{+0.0073}_{-0.0056}$ & $0.053\pm0.088$ & $28.549^{+0.212}_{-0.129}$ & $0.08440\pm0.00038$\\
    TRAPPIST-1 b & $2566\pm26$ & $5.2395^{+0.0073}_{-0.0056}$ & $0.053\pm0.088$ & $20.83\pm0.155$ & $0.08590\pm0.00037$ \\
    LTT 1445 A b & $3340\pm150$ & $4.982^{+0.040}_{-0.065}$ & $-0.34\pm0.09$ & $30.2\pm1.7$ & $0.0454\pm0.0012$\\
    GJ 1132 b & $3229^{+78}_{-62}$  & $5.037^{+0.034}_{-0.026}$ &$-0.17\pm0.15$ & $15.26^{+0.59}_{-0.45}$ & $0.04943 \pm 0.00015$\\
    GJ 486 b & $3317^{+36}_{-37}$ & $4.9111^{+0.0068}_{-0.0110}$  & $-0.15^{+0.13}_{-0.12}$ & $11.380^{+0.074}_{-0.150}$ & $0.037244^{+0.000059}_{-0.000056}$ \\ 
    LHS 3844 b & $3036\pm77$ & $5.06\pm0.01$ & $0\pm0.5^{a}$ & $7.109\pm0.029$ & $0.0635\pm0.0009$ \\
    GJ 1252 b & $3458\pm157$ & $4.83497\pm0.00292$ & $0.1\pm0.1$ & $5.03\pm0.27$ & $0.0277\pm0.0011$ \\
    TOI-1685 b & $3575\pm53$ & $4.83^{+0.043}_{-0.039}$ & $0.3\pm0.1$ & $5.46\pm0.08$ & $0.027494^{+0.000547}_{-0.000531}$ \\
    GJ 367 b & $3522\pm70$ & $4.776\pm0.026$ & $-0.01\pm0.12$ & $3.329\pm0.085$ & $0.01399\pm0.00028$ \\
\enddata
\tablecomments{Stellar and system parameters used for nested sampling, in order of irradiation temperature. Values are taken from: TRAPPIST-1~c/b \citep{agol21}, LTT~1445~A~b \citep{Wachiraphan24}, GJ~1132~b \citep{xue24}, GJ~486~b \citep{mansfield24}, LHS~3844~b \citep{Vander19}, GJ~1252~b \citep{Crossfield22}, TOI-1685~b (\citealt{burt2024}, $a/R_{\star}$ and $R_{p}/R_{\star}$ from \citealt{luque24}), and GJ~367~b \citep{goffo2023}.$^{a}$This value has not been measured and we assume a relatively unconstrained prior. }\label{tab:hoststars}
\end{deluxetable*}

How are dayside temperatures determined through observations? Secondary eclipse observations can help to constrain the planet's dayside \textit{brightness temperature}, which is the best-fit temperature of a blackbody emitter in a given wavelength band.  This value can differ slightly from the true dayside temperature, since observations do not span the full wavelength range of the planet's thermal emission and are thus affected by wavelength-dependent gaseous molecular absorption or emissivity of the surface.

Brightness temperatures are derived from planet-to-star contrast ratios during secondary eclipse (eclipse depths) and require taking into account uncertainty in orbital, planetary, and stellar parameters. Individual observations, however, have accomplished this by using slightly different methodology. This includes the use of spatially resolved (e.g., \citealt{zieba23,ducrot23}) vs. 0-D energy balance models (e.g., \citealt{xue24,mansfield24}), as well as fitting for per-frequency flux (e.g., \citealt{greene23}) vs. planet-to-star contrast ratios.  %Differences in stellar models can have a similar effect.

In this work, we combine the reported eclipse depths and uncertainties from JWST and Spitzer emission observations of TRAPPIST-1~c \citep{zieba23}, TRAPPIST-1~b \citep{greene23,ducrot23}, LTT~1445~A~b \citep{Wachiraphan24}, GJ~1132~b \citep{xue24}, GJ~486~b \citep{mansfield24}, LHS~3844~b \citep{kreidberg19}, GJ~1252~b \citep{Crossfield22}, TOI-1685~b \citep{luque24}, and GJ~367~b \citep{zhang24} to homogeneously re-derive dayside brightness temperature ratios.  These observations utilize five instruments that have different wavelength coverage, including: the Spitzer Infrared Array Camera (IRAC, \citealt{fazio2004}) Channel 2 ($4-\SI{5}{\micron}$), the JWST Near Infrared Spectrograph (NIRSpec, \citealt{jakobsen2022near}) G395H grating ($2.9-\SI{5.1}{\micron}$), the MIRI \citep{rieke2015} Low-Resolution Spectrometer (LRS, $5-\SI{12}{\micron}$), and MIRI imaging filters F1280W ($11.6-\SI{14.0}{\micron}$) and F1500W ($13.5-\SI{16.5}{\micron}$).

To remove potential biases and avoid differences in modeling schemes used in deriving brightness temperature ratios, we use a nested sampling approach (with \texttt{dynesty}, \citealt{speagle20}) that takes into account uncertainties in stellar effective temperature, stellar surface gravity [$\log{(g)}$], stellar metallicity ([M]), as well as the orbital semi-major axis to stellar radius ratio ($a/R_{\star}$) and planet-to-star radius ratio ($R_{p}/R_{\star}$).  These parameters are shown in Table \ref{tab:hoststars}.  Following \citet{xue24}, the broadband planet-to-star flux ratio ($F_{p}/F_{\star}$) is determined assuming an isothermal blackbody emitter for the planet,

\begin{equation} \label{eq:FpFs}
    \frac{F_{p}}{F_{\star}}=\left(\frac{R_{p}}{R_{\star}}\right)^{2}\times
    \frac{\int\frac{\pi B_{\lambda}(\mathcal{R}\times T_{d,max}(T_{\star},a/R_{\star}))}{hc/\lambda}W_{inst,\lambda}d\lambda}{\int\frac{M_{\lambda}(T_{\star},\log{(g)},[\mathrm{M}])}{hc/\lambda}W_{inst,\lambda}d\lambda},
\end{equation}
where $B_{\lambda}$ is the Planck function and $M_{\lambda}$ is the  model stellar flux. $W_{inst,\lambda}$ is the throughput (photon to electron conversion efficiency) function for each instrument, which we determine from \texttt{Pandeia~4.0} \citep{pontoppidan16}. After integrating over the instrument bandpass, the residual of this value (compared to the reported eclipse depth) is then used to calculate the log evidence during nested sampling, which gives a posterior distribution for the brightness temperature ratio $\mathcal{R}$.

The choice of stellar model has a large impact on the modeled emission spectra of rocky planets. Stellar models are not perfect and often conflict with observed stellar spectra and as estimates of $\mathcal{R}$ inherently require accurate stellar models, this can lead to inaccuracies in the interpretation of eclipse spectra.  However, many observations have used a single stellar model to interpret eclipse data (e.g., \citealt{kreidberg19,zieba23,luque24,Wachiraphan24}). Here, we use both the SPHINX M-dwarf Spectral Grid (\citealt{iyer23,iyer_2023_zenodo}, updated May 30 2024), which has been extensively vetted through observations of M dwarfs, and PHOENIX stellar models \citep{husser13}. The SPHINX models include improvements over PHOENIX models specific to M type stars, such as updated molecular line lists \citep{iyer23}. While no stellar model is perfect, SPHINX models have generally performed better than or comparable to PHOENIX models in terms of fitting (absolute) stellar flux in the mid-infrared from JWST results (\citealt{ih23}, also JWST GO Program 1846 and 3730 teams, private communication).  However, both models show disagreement with observed spectra in the visible-near-infrared (e.g., \citealt{lim23,diamond24,radica2025}), and as there is not yet a population-level study of the accuracy of SPHINX vs. PHOENIX models in JWST M dwarf observations, we include both. Brightness temperature ratios can vary significantly between the two models (up to $\sim4\%$ or $\sim1\,\sigma$), highlighting the importance of accounting for uncertainties in stellar modeling.  For nested sampling, we use a linear interpolation scheme for both SPHINX\footnote{\url{https://github.com/ideasrule/sphinx}} and PHOENIX (using \texttt{pysynphot}, \citealt{pysynphot}) to estimate model spectra between grid points.

\subsubsection{Broadband Versus Spectral Fitting for Determining Brightness Temperatures}

Five of the nine planets analyzed in this study involve spectroscopically-resolved data, which involves an additional uncertainty that we now discuss. With JWST, it is now possible to retrieve wavelength-resolved eclipse depths for rocky planets (eclipse spectra), giving information on possible atmospheric and surface spectral features while simultaneously constraining the dayside temperature.  Assuming a blackbody, such spectra can be fitted in $F_{p}/F_{\star}$ vs. wavelength space to find the best-fit blackbody temperature.

On the other hand, early observations suggest that using broadband `white-light' curves are more effective at minimizing the effects of systematics and red noise, leading to more robust \textit{bandpass-integrated} eclipse depths. Spectroscopic fitting may also be more sensitive to outlier points and the exact shape of the blackbody spectrum. 
However, the systematic ramps of LRS are likely wavelength-dependent \citep{zhang24,xue24,mansfield24}, and the `white-light' approach cannot account for such effects.  

Despite using the same fundamental data, these techniques yield different values for $\mathcal{R}$.  For GJ~486~b, fitting the spectrum lowers the derived $\mathcal{R}$ value by $\sim5\%$ ($\sim 3\,\sigma$) when compared to white-light, greatly affecting interpretation of the data.  We find that a synthetic reconstruction of the white-light eclipse depth using the eclipse spectrum is also lower than the observed white-light eclipse depth for GJ~1132~b, GJ~367~b, and TOI-1685~b. Future M-Earth observations with LRS will be needed to better understand the impact of wavelength-dependent systematics on derived brightness temperatures, as this may serve as an additional false positive for atmosphere detection.  

In this work, we mainly adopt brightness temperature ratios using `white-light' broadband eclipse depths, although we report values from spectral fitting as well.  For spectral fits, our nested sampling algorithm uses the same methodology as Eq. \ref{eq:FpFs}, but over multiple wavelength bins simultaneously.  For these fits, we use the LRS emission spectra reported in \citet{zhang24,xue24,mansfield24,Wachiraphan24} and the NIRSpec G395H prayer-bead eclipse spectrum reported in \citet{luque24}.  Our homogeneously re-derived brightness temperature ratios are shown in Table \ref{tab:Rvalues}, and are consistent with those derived in the original observations (see Appendix Table \ref{tab:comparison}).

\begin{deluxetable*}{lcccccccc} 
\tablecaption{Planet Parameters and Homogeneously-Derived $\mathcal{R}$ Values Used in This Study}
\tablewidth{0pt}
\tablehead{
\colhead{Planet} & \colhead{$T_{irr}$} & \colhead{Eclipse Depth(s)} & \multicolumn{2}{c}{$\mathcal{R}$} & 
\colhead{Instrument(s)} & \colhead{$M_p$} & \colhead{$R_p$} \\[-2mm]
\colhead{} & \colhead{(K)} & \colhead{(ppm)} & \colhead{SPHINX} & \colhead{PHOENIX} & \colhead{} & \colhead{$(M_{\oplus})$} & \colhead{$(R_{\oplus})$}}

\startdata
    TRAPPIST-1 c & $480\pm5$ & $421\pm94$ & $0.877^{+0.073}_{-0.075}$ & $0.903^{+0.075}_{-0.082}$ & MIRI F1500W  & $1.31\pm0.06$ & $1.10\pm0.01$ \\ \hline
    TRAPPIST-1 b$^{a}$ & $562\pm5$ & $452\pm86, 775\pm90$  & $0.910^{+0.037}_{-0.036}$ & $0.933^{+0.039}_{-0.040}$& MIRI F1280W, F1500W & $1.37\pm0.07$ & $1.12\pm0.01$ \\ 
    Greene+23 &  & $861\pm99$  & $0.993^{+0.049}_{-0.052}$ & $1.021^{+0.054}_{-0.055}$ & F1500W & & \\ \hline
    LTT 1445 A b & $600\pm30$ & $41\pm9$ & $0.950^{+0.063}_{-0.071}$ & $0.955^{+0.066}_{-0.072}$ & MIRI LRS & $2.7\pm0.2$ & $1.3\pm0.1$  \\ 
    Spectral Fit & & & $0.948^{+0.043}_{-0.043}$ & $0.954^{+0.048}_{-0.046}$  & & & \\ \hline
    GJ 1132 b & $826\pm14$ & $140\pm17$ & $0.940^{+0.043}_{-0.040}$ & $0.952^{+0.042}_{-0.044}$& MIRI LRS & $1.84\pm0.19$ & $1.19\pm0.04$\\
    Spectral Fit & & & $0.902^{+0.038}_{-0.038}$ & $0.914^{+0.038}_{-0.038}$ & & & \\ \hline
    GJ 486 b & $985\pm10$ & $135.5\pm4.9$ &  $0.973^{+0.016}_{-0.017}$ & $0.978^{+0.016}_{-0.015}$& MIRI LRS & $2.77\pm0.07$ & $1.29\pm0.02$  \\
    Spectral Fit & & & $0.922^{+0.016}_{-0.015}$ & $0.932^{+0.013}_{-0.014}$ & & & \\ \hline
    LHS 3844 b & $1138\pm28$ & $380\pm40$ & $0.996^{+0.033}_{-0.034}$ & $1.002^{+0.033}_{-0.034}$& IRAC Channel 2 &   $2.2\pm1.0^{b}$ & $1.30\pm0.02$  \\ \hline
    GJ 1252 b & $1540\pm98$ & $149^{+{25}}_{-32}$ & $1.067^{+0.094}_{-0.105}$ & $1.035^{+0.090}_{-0.103}$& IRAC Channel 2 & $1.32\pm0.28$ &  $1.18\pm0.08$ &\\ \hline
    TOI-1685 b${^a}$ & $1541\pm40$ & $119^{+23}_{-18}$ & $1.066^{+0.080}_{-0.069}$ & $1.008^{+0.076}_{-0.058}$ & NIRSpec G395H NRS 2 & $3.03\pm0.33$ & $1.38\pm0.04$ \\ 
    Spectral Fit & & & $0.991^{+0.035}_{-0.039}$ & $0.976^{+0.033}_{-0.035}$ & NIRSpec G395H & & & \\ \hline
    GJ 367 b & $1930\pm45$ & $79\pm4$ & $1.074^{+0.047}_{-0.047}$ & $1.035^{+0.040}_{-0.041}$ & MIRI LRS & $0.63\pm0.05$ & $0.70\pm0.02$\\
    Spectral Fit$^{a}$ & & & $1.002^{+0.049}_{-0.045}$ & $0.966^{+0.044}_{-0.039}$ & & & \\
\enddata
\tablecomments{LHS~1478~b data are not included in our sample due to issues discussed in Appendix \ref{sec:ap_lhs}. Planetary parameters and eclipse depths are from: TRAPPIST-1~c \citep{agol21,zieba23}, TRAPPIST-1~b \citep{agol21,greene23,ducrot23}, LTT~1445~A~b \citep{Wachiraphan24}, LHS~3844~b \citep{Vander19,kreidberg19}, GJ~1132~b \citep{xue24}, GJ~486~b \citep{mansfield24},
GJ~1252~b \citep{Shporer20,Crossfield22}, TOI-1685~b \citep{burt2024,luque24}, and GJ~367~b \citep{goffo2023,zhang24}. $^{a}$See Appendix \ref{ap:data_considerations} for a discussion of data considerations for these planets. $^{b}$LHS~3844~b does not have a measured mass and we adopt the unconstrained value assumed in \citet{diamond21}.
}\label{tab:Rvalues}
\end{deluxetable*}

\begin{figure*}
    \centering
    \gridline{
        \fig{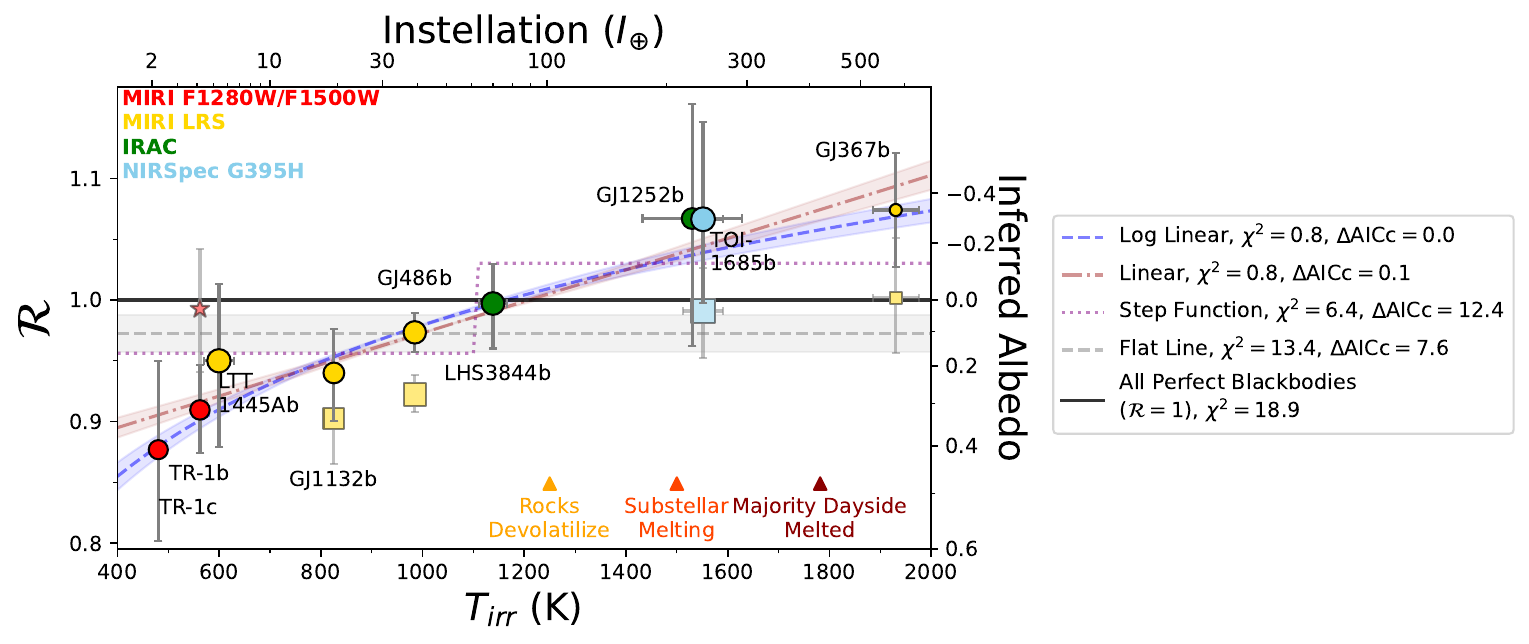}{0.8\linewidth}{(a) SPHINX}     
    }
    \gridline{
    \fig{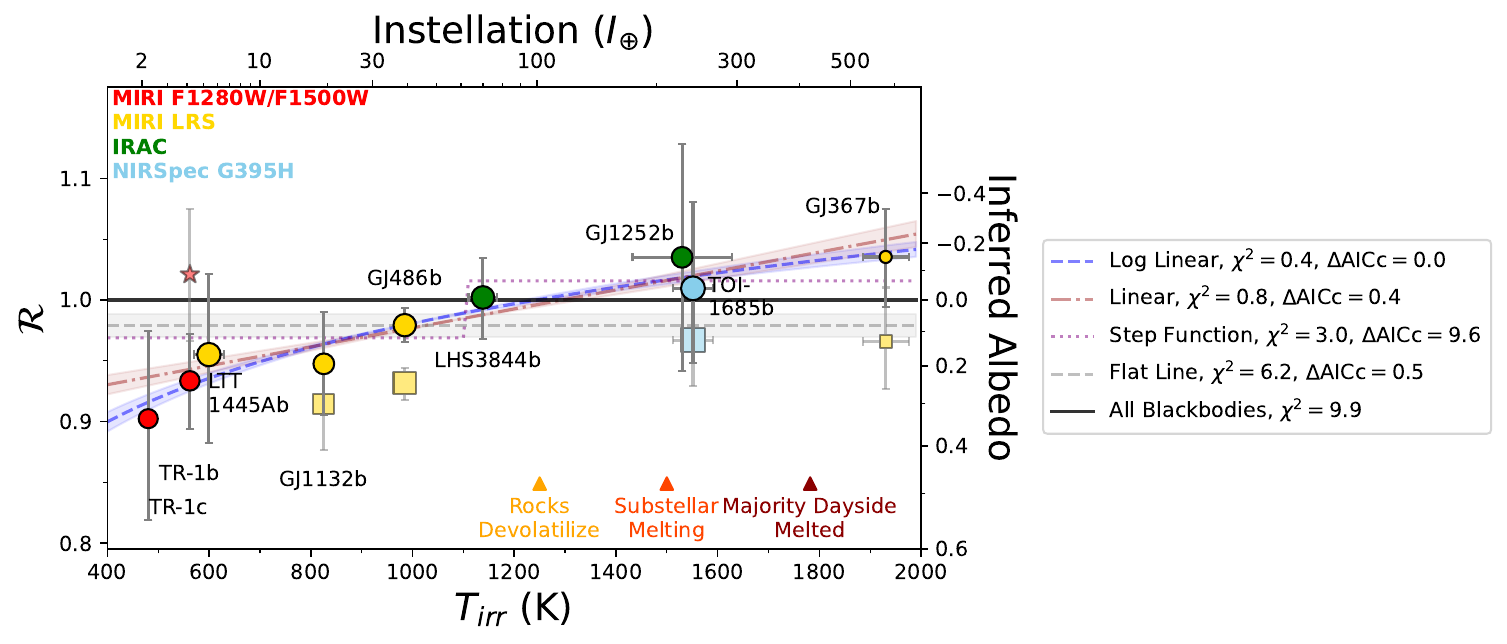}{0.8\linewidth}{(b) PHOENIX}
    }
    \caption{
    Our homogeneously-derived brightness temperature ratios $\mathcal{R}$ (the measured dayside brightness temperature compared that of a perfect blackbody) using (a) SPHINX and (b) PHOENIX stellar models as a function of irradiation temperature for M-Earths with emission data (Table \ref{tab:Rvalues}).  Point radii are proportional to planet radii.
   There is an apparent trend in brightness temperature as a function of irradiation temperature; $\mathcal{R}$ increases with increasing temperatures. We include the chi-square statistic $\chi^{2}$ and $\Delta$AICc (corrected Akaike Information Criterion) for five possible simple functions describing the observed data, noting that a trend is strongly favored over a flat line using SPHINX stellar models. `white-light' broadband data (values used in our study) are shown as circles, whereas fits to spectra are shown in squares.  The star represents TRAPPIST-1~b F1500W data originally presented in \citet{greene23} not used in this study (see Appendix  \ref{sec:ap_trappist}).  Rocky surfaces of composition similar to Earth's are expected to largely devolatilize around $\sim\SI{1250}{K}$ and completely melt by \SI{1500}{K} \citep{lutgens00,mansfield19}.}
    \label{fig:trend}
\end{figure*}

\subsection{Surface Modeling} 
\label{sec:calc_albedo}

To estimate the effects of surface compositions on brightness temperatures, we largely follow the simplified 0-D energy balance model of \citet{mansfield19}. Diverging from \citet{mansfield19}, we use the approximations presented in the \citet{hapke12} scattering model and single scattering albedo ($w$) data derived in \citet{hu12} to determine the spherical reflectance $r_{s}$ (equivalent to spherical albedo in an isotropic scatterer):

\begin{equation}
    r_{s}(\lambda)=r_{0}(\lambda)\left(1-\frac{1-r_{0}(\lambda)}{6}\right),
\end{equation}
where $r_{0}$ is the `diffusive reflectance':
\begin{equation}
    r_{0}(\lambda)=\frac{1-\sqrt{1-w(\lambda)}}{1+\sqrt{1-w(\lambda)}}.
\end{equation}
Hemispherical emissivity $\varepsilon_{h}$ is determined via,

\begin{equation}
    \varepsilon_{h}(\lambda)=1-r_{s}(\lambda).
\end{equation}

Similarly, we use the \citet{hapke12} model to derive the geometric albedo $A_{g}$, which controls the amount of light backscattered towards the observer (reflected light), ignoring the opposition surge effect:
\begin{equation}
    A_{g}(\lambda)=0.49r_{0}(\lambda)+0.196 r_{0}^{2}(\lambda).
\end{equation}
The strength of the opposition surge effect is highly dependent on the surface material composition and grain size (e.g., \citealt{jost16}) and it not calculable \textit{a priori}.  Regardless, it has little effect on the eclipse spectrum due to being a purely observational reflected light effect that does not affect energy balance. We note that the distinction between spherical (wavelength-dependent Bond) albedo and geometric albedo was erroneously not made in \citet{mansfield19}, and discuss implications in Section \ref{sec:false_pos}.

\section{Results}\label{sec:results}

\subsection{Trend in Brightness Temperature Ratio with Irradiation Temperature} 

We compare our homogeneously re-derived dayside brightness temperatures ratios (Table \ref{tab:Rvalues}) as a function of irradiation temperature in Fig. \ref{fig:trend}.  While all individual observations are consistent with a no thick atmosphere null hypothesis, together these results show a trend in brightness temperature as a function of irradiation temperature; $\mathcal{R}$ decreases with colder temperatures. %Planets colder than GJ 486b are generally less consistent with a blackbody emitter. 

To statistically evaluate this trend, we use both goodness-of-fit hypothesis testing and model comparison based on information-theory criteria.  We adopt the `flat line' (i.e., no trend) model as our null hypothesis to compare to two simple linear models with Student's t-tests.  These linear models consist of a standard linear regression,
\begin{equation}
    \mathcal{R}=\alpha_{0}+\alpha_{1}T_{irr},
\end{equation}
with weighted least-squares fit coefficients (using the SPHINX model-derived $\mathcal{R}$ values) of $\alpha_{0}=0.8431\pm0.0132$ and $\alpha_{1}=0.000130\pm0.000013$,
and a `log-linear' regression,
\begin{equation}
    \mathcal{R}=\beta_{0}+\beta_{1}\ln(T_{irr}),
\end{equation}
with coefficients $\beta_{0}=0.0413\pm0.0896$, $\beta_{1}=0.1358\pm0.0130$.  Through $t$-tests, we reject the null hypotheses that $\alpha_{1}=0$ or $\beta_{1}=0$ (i.e., no trend) at $p$-values of $1.8\times10^{-5}$ ($4.3\,\sigma$) and $1.6\times10^{-5}$ ($4.3\,\sigma$), respectively.

To further compare various simple models, we calculate the chi-square statistic ($\chi^{2}$), relative corrected Akaike Information Criterion (AICc), and Bayesian Information Criterion (BIC) values, shown in Table \ref{tab:stats}. %Simple metrics like \textit{reduced} chi-square and BIC can be misleading for small sample sizes, as the expected value can itself have a large variance.  
AICc is specifically used for small sample sizes by penalizing more complex models \citep{anderson2002}.  According to all of our metrics, the log-linear description of the data is favored over the null hypothesis. The log-linear model finds the best AICc, with a $\Delta\mathrm{AICc}=7.6$ over the flat line model, which implies `considerably less' support for the latter \citep{anderson2002}.  The $\Delta$BIC value (10.2) also surpasses the threshold indicating `very strong evidence' in favor of the log-linear model ($\Delta\mathrm{BIC}=10$, \citealt{raftery1995}), although this value may be overly optimistic for our small sample size.  This trend suggests a $T_{irr}$-dependent process(es) that increases the inferred albedo of colder planets. 

However, nested sampling using PHOENIX stellar models tend to push $\mathcal{R}$ towards unity (Fig. \ref{fig:trend}).  In this case, the evidence for a log-linear trend, with coefficients $\beta_{0}=0.3701\pm0.0609$ and $\beta_{1}=0.0884\pm0.0088$, or a linear trend where
$\alpha_{0}=0.8991\pm0.0126$ and $\alpha_{1}=(7.79\pm1.17)\times10^{-5}$,
is much weaker.  While we still reject the null hypotheses that $\alpha_{1}=0$ or $\beta_{1}=0$ with $p$-values of $2.9\times10^{-4}$ ($3.6\,\sigma$) and $2.6\times10^{-5}$ ($4.2\,\sigma$), respectively, $\Delta$AICc suggests that the flat line, linear, and log-linear models are roughly equally as likely.  Similarly, the lower $\mathcal{R}$ values found from spectral fitting significantly reduce the statistical confidence of the trend, with a flat line being marginally preferred in the PHOENIX spectral fitting case (Table \ref{tab:stats}).  Thus, while SPHINX (which is a stellar grid specifically dedicated for the low-mass stars in this study) fitting using broadband eclipse depths shows strong evidence for a trend in $\mathcal{R}$, we emphasize that this is a tentative identification that requires more data, more precise stellar modeling, or a better understanding of the effects of wavelength-dependent systematics on shallow eclipse spectra to further support.

\begin{deluxetable}{lcccccccc}
\tablecaption{Statistics of Simple Functions Describing the Data}
\tablewidth{0pt}
\label{tab:stats}
\tablehead{
\colhead{Function} & \colhead{$\chi^{2}\,(N=9)$} & \colhead{$\Delta$AICc} & 
\colhead{$\Delta$BIC}}
\startdata
    \multicolumn{4}{c}{SPHINX Data} \\ \hline
    Log-Linear & 0.8  & 0 & 0 \\
    Linear & 0.8  & 0.1 & 0.1 \\
    Step Function & 5.2 &  11.5 & 6.5 \\
    Flat Line & 13.4 & 7.6 & 10.2  \\
    All Perfect Blackbodies ($\mathcal{R}=1$) & 18.9 & -  & - \\ \hline
    \multicolumn{4}{c}{PHOENIX Data} \\ \hline
    Log-Linear & 0.4  & 0 & 0 \\
    Linear & 0.8 & 0.4 & 0.4 \\
    Step Function & 2.6 & 9.1 & 4.1 \\
    Flat Line & 6.2 & 0.5 & 3.1  \\
    All Perfect Blackbodies ($\mathcal{R}=1$) & 9.9 & - & - \\ \hline
    \multicolumn{4}{c}{SPHINX Data (Spectral)} \\ \hline
    Log-Linear & 4.6  & 0.6 & 1.8 \\
    Linear & 3.9  & 0 & 1.2 \\
    Step Function & 1.3 &  3.8 & 0  \\
    Flat Line & 9.5 & 2.0 & 5.8  \\
    All Perfect Blackbodies ($\mathcal{R}=1$) & 40.1 & -  & - \\ \hline
    \multicolumn{4}{c}{PHOENIX Data (Spectral)} \\ \hline
    Log-Linear & 4.8  & 3.2 & 0.8 \\
    Linear & 4.6  & 3.4 & 0.6 \\
    Step Function & 1.5 &  7.8 & 0.2  \\
    Flat Line & 6.5 & 0 & 0  \\
    All Perfect Blackbodies ($\mathcal{R}=1$) & 41.3 & -  & - \\ \hline
\enddata
%\tablecomments{}
\end{deluxetable}

\subsection{Effects of Surface Composition on Brightness Temperature} \label{sec:surface}

All planets in this study lack evidence for a thick atmosphere. In the absence of an atmosphere, infrared emission observations  probe the composition of rocky exoplanet surfaces. Here, we explore possible \textit{geophysical} processes that could explain a trend in $\mathcal{R}$, making hotter solid surfaces darker.

\subsubsection{Space Weathering}

Space weathering is a generalized term that refers to surface alteration primarily by stellar winds and micrometeorite impacts.  The surfaces of Mercury and the Moon are darker than pulverized rocks of similar composition \citep{hapke01}, having low estimated Bond albedos of 0.06 and 0.13, respectively \citep{mallama2002,matthews2008}. This darkening is also important for asteroids (e.g., \citealt{chapman04,pieters16}).  Darkening occurs primarily from the conversion of iron (Fe) locked in silicates on the surface to nm-sized nanophase metallic Fe (npFe$^{0}$) and larger-grained microphase Fe (mpFe$^{0}$). These particles darken a thin ($\lesssim\mu$m) surface layer, lowering planetary albedo. In the case of an Fe-poor, carbon-rich surface such as that of Mercury \citep{pieters16}, darkening is thought to be due to graphite, possibly a relict of a graphite flotation crust \citep{keppler19}.   Space weathering typically reddens reflectance spectra by reducing albedo primarily in the visible-NIR for npFe$^{0}$, or reduces albedo over all wavelengths in the case of mpFe$^{0}$ or  graphite-coating \citep{pieters16}.  The effects of space weathering on remote sensing observations are so ubiquitous that the spectral slope (reddening) induced by npFe$^{0}$ contamination can be used to estimate the exposure ages of asteroids (e.g., \citealt{jedicke2004,willman2011,marchi2012}). Space weathering can be prevented by even thin, Mars-like atmospheres. 

Most previous theoretical studies of exoplanet surfaces (e.g., \citealt{hu12,mansfield19,whittaker22,hammond24}) model the emission spectra of \textit{fresh}, unweathered regolith and thus likely overestimate the impact of composition on brightness temperature for space-weathered planets.  The instellation (and stellar type) dependence of space weathering is not well understood. \citet{zieba23} used stellar wind strength scaling to estimate a \textit{space weathering timescale} for TRAPPIST-1 c (the least irradiated planet considered here) of $10^{2}-10^{3}$ yr, compared to $10^{5}- 10^{7}$ yr for the Moon (according to \citealt{hapke77}), as stellar winds are expected to be more intense for low-mass stars \citep{johnstone2015}.  If this is correct, airless M-Earths would have to experience dayside resurfacing through volcanism on very short timescales to produce detectable nonzero surface albedo. Indeed, the Spitzer phase curve of LHS~3844~b has suggested that it is highly space weathered \citep{lyu2024}. If space weathering is efficient and atmospheres are lacking on M-Earths, then very low-albedo surfaces are likely on these planets. 

To quantify the effects of space weathering on measured brightness temperatures, we simulate the effects of the mixture of a host material and npFe$^{0}$ or graphite absorbing particles, following \citet{hapke01} and \citet{lyu2024}, using single scattering albedo profiles from \citet{hu12}. \citet{lyu2024} found that the phase curve of LHS~3844~b is most consistent with either mixtures of either 5 wt\% npFe$^{0}$ or 5 wt\% graphite.  They consider 5\% an upper limit given the $\sim5$ wt\% Fe crustal content of the Earth and the Moon \citep{lucey95,taylor01,taylor06}.
This is noticeably higher than contamination in lunar soil samples, which suggest $\sim0.1-0.5\,\mathrm{wt}\%$ npFe$^0$ \citep{hapke01,noble2007experimental}, further supporting efficient space weathering for M-Earths. 

\begin{figure*}
    \centering
    \includegraphics[width=0.9\linewidth]{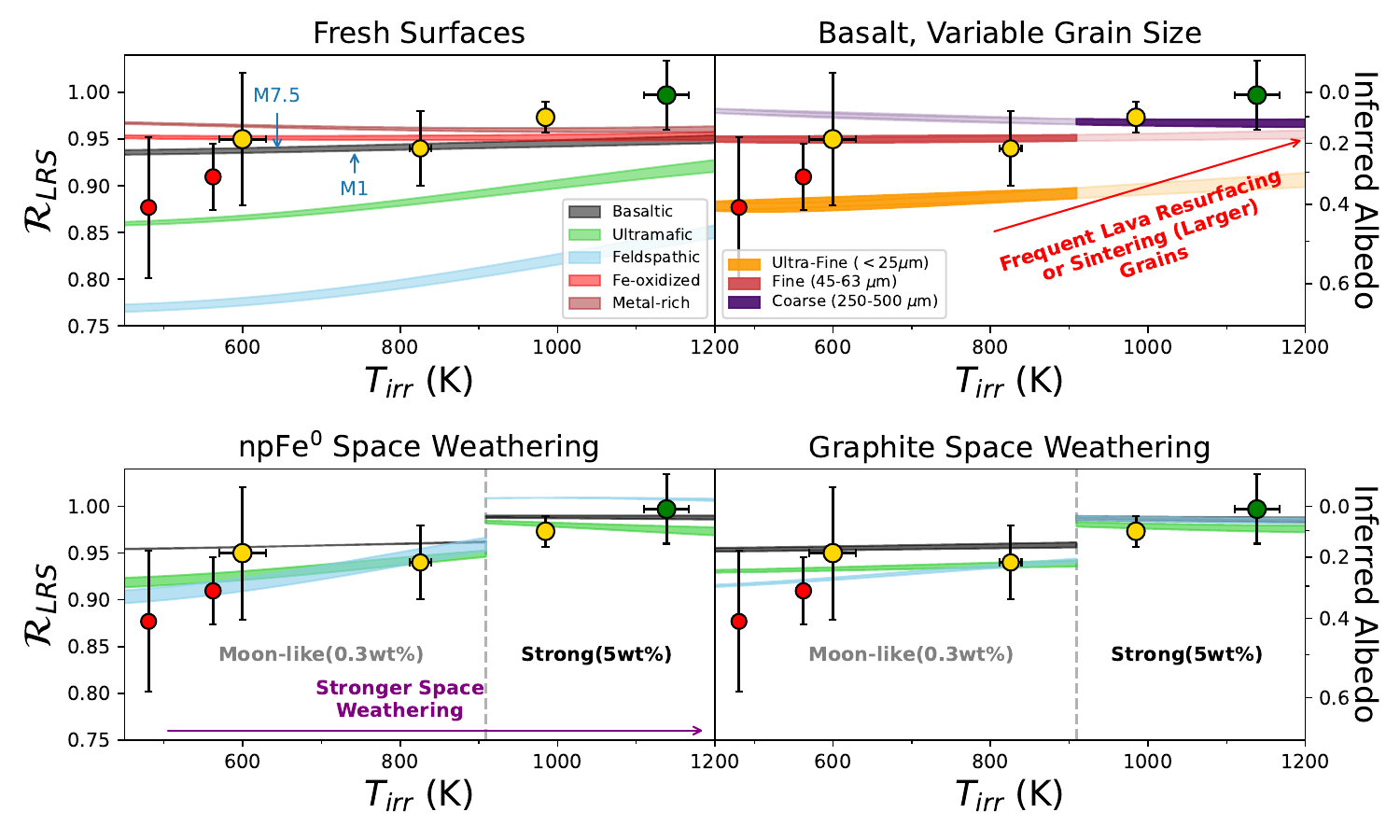}
    \caption{Simulated effects of `bare-rock' fresh (unweathered) surface albedo profiles from \citet{hu12}, varying regolith grain sizes of basalt, and npFe$^{0}$/graphite space weathering on simulated emission observations in the MIRI LRS bandpass ($5-\SI{12}{\micron}$). %Due to a lack of strong absorption features in these ranges, 
    Brightness temperature ratios in the MIRI F1500W and Spitzer IRAC2 bandpasses show similar trends. Planets too hot for solid surfaces ($T_{irr}\gtrsim\SI{1250}{K}$) are not shown. Results are dependent on the exact stellar properties, and we include results (using SPHINX stellar models) for a GJ~367-like M1 star and a TRAPPIST-1-like M7.5 star as bounding cases defining the width of each colored band. This roughly encompasses the spread expected for M-Earths. A trend of increasing brightness temperature ratio ($\mathcal{R}$) with irradiation temperature ($T_{irr}$) can be explained by grain sizes increasing with temperature or stronger space weathering on closer-in planets, as explored in Section \ref{sec:surface}. As in Fig. \ref{fig:trend}, point colors represent the instrument from which $\mathcal{R}$ is derived: red (MIRI photometry), yellow (LRS), and green (IRAC Channel 2).
    }
    \label{fig:spaceweather}
\end{figure*}

Due to uncertainties in the exact behavior of space weathering around M stars, we test the effects of both weak Moon-like ($0.3\,\mathrm{wt}\%$ npFe$^{0}$ or graphite) and stronger LHS~3844~b-like ($5\,\mathrm{wt}\%$) space weathering on the albedo profiles of \citet{hu12}. We simulate MIRI LRS emission observations for a wide range of irradiation temperature and surface types. We find (Fig. \ref{fig:spaceweather}) that universal LHS~3844~b-like highly space weathered surfaces, predicted by stellar wind scaling arguments, \textit{cannot} explain a trend in brightness temperature alone, as strong space weathering pushes $\mathcal{R}$ very close to unity. However, if space weathering is much weaker for rocky planets around M-stars than inferred in \citet{zieba23}, moderate-albedo surfaces (e.g., fresh basalts or ultramafic grains) that are more space-weathered on close-in planets are a possible explanation for the observed trend.  We note that, while npFe$^{0}$ is the major albedo-altering space weathering contaminant for the Moon and most asteroids \citep{pieters16,denevi2023}, that this may not be the case for all M-Earths.  The expected surface mineralogy of M-Earths is largely unconstrained, and iron- and carbon-poor surfaces may be subject to different forms of space weathering not yet understood.

\subsubsection{Regolith Grain Size}

The albedo profiles used in previous surface studies \citep{hu12,mansfield19,lyu2024} assume a fine-grained regolith formed by long-term weathering of surface rock. Regolith formation on planet-sized airless bodies is generally thought to be governed by `surface gardening' processes spurred by micrometeorite and solar wind bombardment \citep{melosh1989}. However, grains can be coarsened by solid-state deformation when close to the melting point (sintering, e.g., \citealt{demirci2017,mergny2024}), or by volcanic resurfacing.

Larger grain sizes decrease visible and near-infrared reflectance of particulate regolith (e.g., \citealt{zaini12,zhuang23}), both decreasing Bond albedo and the magnitude of spectral features. %Natural slab surfaces that have not experienced widespread weathering show reflectance similar to that of large-sized ($\gtrsim\SI{675}{\micron}$) grains \citep{zhuang23}.  
Lunar regolith is considered fine-grained, with bulk optical properties dominated by particles $\sim10-\SI{45}{\micron}$ in size \citep{fischer94}.  However, closer-in planets may experience faster volcanic resurfacing \citep{jackson08}. To form the fine-grained regolith that is assumed in the albedo profiles of \citet{hu12} and subsequent works requires long-term weathering that may be reset by lava resurfacing or by high temperature sintering. Basaltic rock begins forming melt glass around $\sim\SI{1250}{K}$ \citep{winter2014principles}, and thus sintering may be important on hotter worlds like GJ~486~b and LHS~3844~b which are near this melting point. Some laboratory experiments suggest this threshold may be near $\sim\SI{1000}{\kelvin}$ for basaltic rock \citep{demirci2017}, however others suggest a higher value of $\sim\SI{1250}{\kelvin}$ for lunar regolith \citep{han2022sintering}. Given the short-term nature of these experiments, it remains uncertain how sintering behaves on multi-million year timescales.

To estimate the effects of grain size on $\mathcal{R}$, we use results from the RELAB Spectral Database\footnote{\url{https://sites.brown.edu/relab/relab-spectral-database/}}. Particulate of the basalt sample 79-3b (the same sample used in the `basaltic' albedo profile of \citealt{hu12}) were crushed and sorted via particle size bins, ranging from $<\SI{25}{\micron}$ to \SI{500}{\micron} (PI: John F. Mustard).  Measured bidirectional reflectance values cover the $0.3-\SI{25.0}{\micron}$ wavelength range. These data were converted to single scattering albedo, spherical reflectance, and hemispherical emissivity values following methods in \citet{hapke12} and \citet{hammond24}.  Results (Fig. \ref{fig:spaceweather}) confirm that grain size can have a significant impact on $\mathcal{R}$, with larger grain sizes leading to universally hotter planets for the same underlying material.  Thus, if hotter planets have higher resurfacing rates and correspondingly coarser surfaces, this can explain a 1-D trend in brightness temperatures.

\subsection{Wavelength-Dependent Effects of Atmospheres on $\mathcal{R}$} \label{sec:atms}

\begin{figure*}
    \centering
    \includegraphics[width=0.85\linewidth]{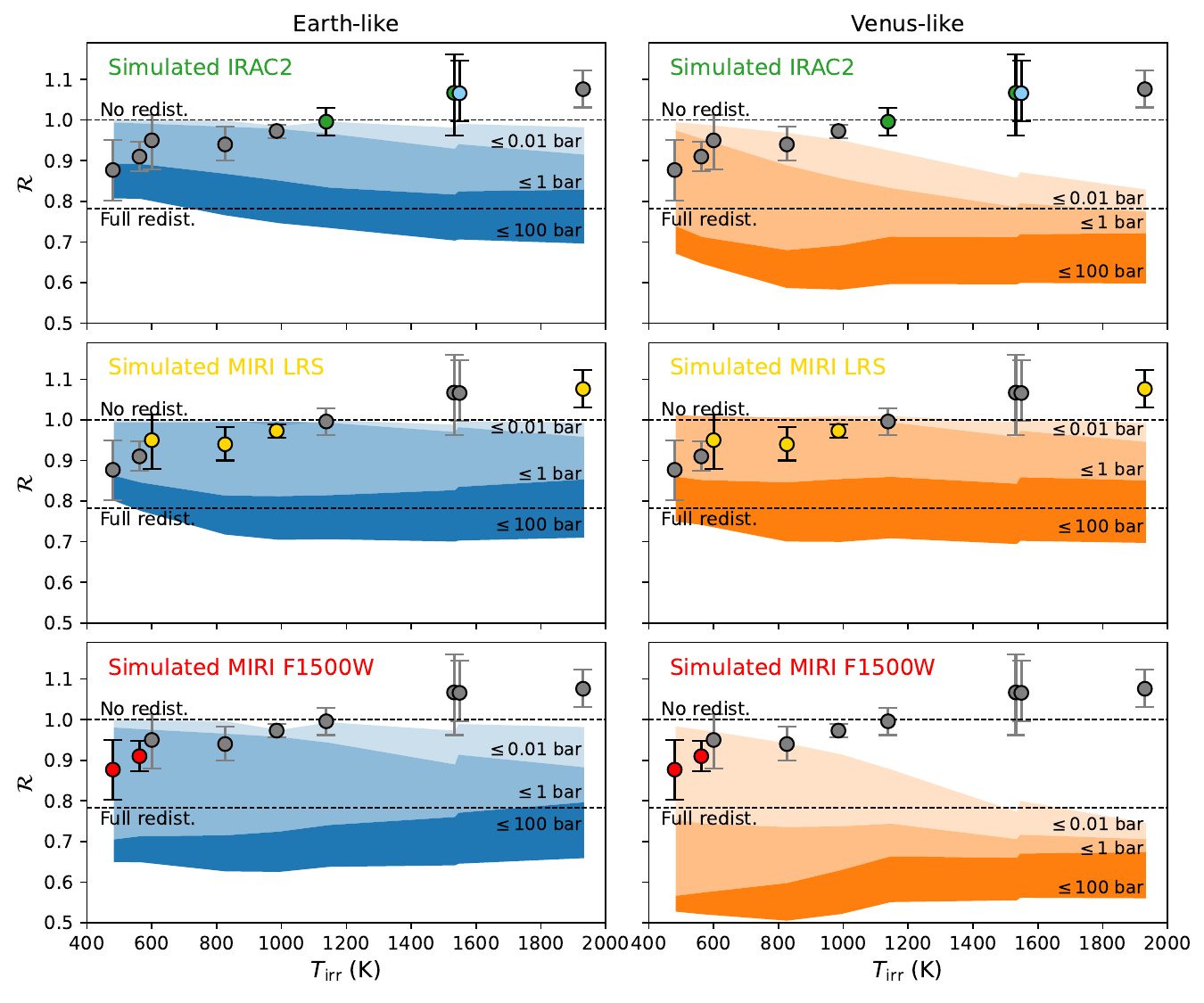}
    \caption{The calculated brightness temperature ratio ($\mathcal{R}$) for each target from atmospheric forward models varying surface pressures, plotted against irradiation temperature.  Each row shows the brightness temperatures calculated in the bandpass of IRAC Channel 2, MIRI LRS, and MIRI F1500W, respectively.  For TOI-1685 b (light blue) the NIRSpec bandpass is used, which most closely overlaps with IRAC Channel 2.  Observations using the respective instrument are highlighted in each panel.  The bottom of each contour corresponds to roughly Mars-like (0.01 bar), Earth-like (1 bar), and Venus-like (100 bar) surface pressures.  The brightness temperature ratio corresponding to an absorber-less atmosphere with no redistribution and full redistribution are shown as dashed lines.  A horizontal offset between GJ 1252 b and TOI-1685 b has been applied for visual clarity.}
    \label{fig:atmosphere}
\end{figure*}

\begin{figure*}
    \centering
    \includegraphics[width=0.75\linewidth]{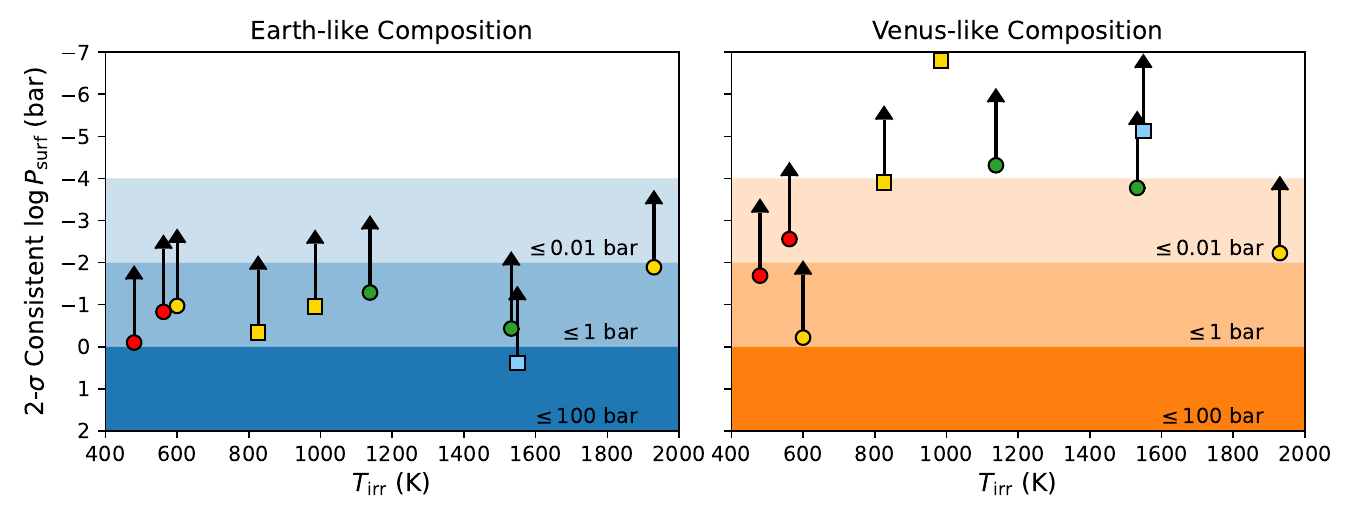}
    \caption{The maximum surface pressure of Earth-like or Venus-like atmospheres consistent with each observation at $2\,\sigma$ as per Fig. \ref{fig:atmosphere}, accounting for each instrument.  Here, the consistent surface pressures are calculated using measured eclipse depths and their uncertainties (rather than $\mathcal{R}$).   The square points indicate that the spectral information available from LRS/G395H was used to find the constraint, where the consistent surface pressures are calculated using goodness-of-fit of the binned eclipse spectra.  The TRAPPIST-1 b point takes both F1280W and F1500W observations into account.}
    \label{fig:psurf}
\end{figure*}

While our sample of planets lack evidence for \textit{thick} atmospheres, we cannot rule out thin/tenuous $<\SI{1}{bar}$ atmospheres because some models predict that volcanic outgassing or volatile replenishment via cometary impacts outpaces atmospheric loss after a $\sim\si{Gyr}$, reviving the atmosphere (e.g., \citealt{kral18,kite20}).  Such revival would be easiest on colder, less irradiated planets. In addition, thin atmospheres are expected to cause negligible heat redistribution to the nightside \citep{koll22}, consistent with phase curve observations \citep{kreidberg19,zhang24,luque24}.  To test the effects that an onset of thin atmospheres would have on observed $\mathcal{R}$ values, we forward model possible cloud-free atmospheres for each target, varying the surface pressure and composition, similar to work done in \citet{whittaker22, ih23, xue24, mansfield24}.  For each system, we use SPHINX stellar models to calculate the stellar spectrum and use the radiative-convective equilibrium code \texttt{HELIOS} \citep{malik17, malik19helios} to calculate the 1D thermal structure and emission spectrum of the planet.  \texttt{HELIOS} employs the scaling relationship developed in \citet{koll22} to calculate the approximate heat redistribution given the surface pressure, equilibrium temperature, and opacity of the atmosphere.  We model the atmosphere for surface pressures from $10^{-4}$ - $10^{2}$ bar in 1-dex intervals and for Earth-like \ce{CO2}-poor (1\% \ce{H2O}, 400 ppm \ce{CO2}) and Venus-like \ce{CO2}-rich (96.5\% \ce{CO2}, 150 ppm \ce{SO2}, 20 ppm \ce{H2O}) compositions \citep{olson18,marcq18}. While this is not representative of the diversity in mantle redox states expected for rocky exoplanets (e.g., \citealt{gaillard2022,LICHTENBERG2025}), we focus on these two scenarios because \ce{N2}/\ce{O2}- and \ce{CO2}-dominated atmospheres are expected to be particularly resilient to x-ray and extreme ultraviolet (XUV)-driven atmospheric loss due to efficient atomic line cooling \citep{tian2009,nakayama22,chatterjee2024}.  We assume a blackbody at the surface.  These forward models are calculated for each planet/star pair based on parameters in Table \ref{tab:hoststars} and \ref{tab:Rvalues} using the median values for stellar effective temperature, metallicity, and gravity as well as planetary radius, mass, and orbital radius.  We use the bandpass of each instrument weighted by the stellar spectrum to calculate the binned eclipse depth, as done in \citet{whittaker22, ih23}.  

We show the brightness temperature ratios $\mathcal{R}$ in each instrument bandpass in Fig. \ref{fig:atmosphere}, for surface pressures corresponding roughly to Venus-, Earth-, and Mars-like surface pressures in 2-dex intervals.  Here, a clear trend in possible atmospheric thickness is tricky to infer because the instrument choice can have a large effect, with bandpasses that targets specific bands, e.g., IRAC Channel~2 or MIRI F1500W for \ce{CO2}, being the most discerning.  However, when viewed in consistent surface pressure space (Fig. \ref{fig:psurf}) upper limits on surface pressure for Venus-like atmospheres become somewhat larger for colder planets, \textit{possibly} indicative of the onset of thin atmospheres. Planets hotter than $T_{irr}\sim\SI{1000}{\kelvin}$ are difficult to reconcile with even thin \ce{CO2}-rich atmospheres.

It is worth considering whether these thin atmospheres are sustainable. Thin ($\lesssim 0.1$ bar) \ce{CO2}-rich atmospheres are likely subject to atmospheric collapse on the nightside for warm tidally-locked M-Earths \citep{wordsworth2015} and thus would require constant resupply via volcanism. In addition, some models predict that atmospheres on M-Earths would require extremely high volcanic outgassing fluxes to balance high thermal escape rates (e.g., \citealt{diamond21,Crossfield22,krissansen2023,diamond24}). For example, \citet{foley2024} estimated that TRAPPIST-1~c would have to have outgassing rates roughly $1-3$ orders of magnitude higher than modern-day Earth to sustain a significant \ce{CO2} or \ce{H2O}-dominated atmosphere, whereas for LHS~3844~b this rises to $3-5$ orders of magnitude due to higher thermal escape fluxes. However, protective magnetic fields \citep{segura10,luo24}, cooling of atmospheres by radiative recombination and atomic line cooling \citep{nakayama22}, and/or potentially high initial volatile inventories \citep{bergin23,pengvalencia24}, could allow M-Earths to retain atmospheres.  Outer, colder M-Earths are more likely to retain some atmosphere because they experience far less atmospheric loss fluxes from processes like thermal escape, solar wind striping, and impact-based erosion (e.g., \citealt{ribas2016,dong18,kite20}). Future observations will be needed to break the degeneracies between thin atmospheres and moderate-albedo surfaces.

Instruments included in this study offer tradeoffs between the efficiency of atmosphere detection and observability of targets. MIRI F1500W can offer efficiently ruling out thick atmospheres with even a small amount of \ce{CO2}, but may be sensitive to false positives due to the limited wavelength coverage \citep{ih23,hammond24}.  False negatives due to thermally-inverted atmospheres have also been suggested for F1500W (\citealt{ducrot23}, also see Section \ref{sec:aerosols}).

On the other hand, MIRI LRS offers the most precise constraints on planetary dayside \textit{effective} temperature and is simultaneously sensitive to select gaseous spectral features, the effects of heat redistribution, and possibly surface mineralogy (e.g., \citealt{whittaker22,first2024,paragas25}). However, detailed characterization incorporating spectral information may only be suitable for very observationally favorable (and hot) targets unlikely to host atmospheres due to low eclipse depths in the MIRI LRS wavelength range for colder targets.  Studying population-level effects of the onset of thin atmospheres should ideally use the same instrument(s), since as shown in Fig. \ref{fig:atmosphere}, the chosen instrument has a large effect on $\mathcal{R}$. Upcoming large-scale eclipse surveys, such as the Hot Rocks Survey \citep{hotrocks} and the Rocky Worlds DDT program 
\citep{Redfield2024}, both slated to use MIRI F1500W, have potential to provide such measurements.

A caveat to our atmosphere analyses is that we assume blackbodies at the surface.  Medium- to high-albedo surfaces on planets with even thin atmospheres are more likely than airless bodies as they are not subject to space weathering, and may modify the planet's energy budget and subsequently $\mathcal{R}$.

\section{Additional Processes That Can Affect Brightness Temperature}\label{ap:additional}
Several other processes not discussed above have the potential to significantly affect $\mathcal{R}$.  Here, we discuss a variety of these processes and their potential impacts on $\mathcal{R}$ for the M-Earths considered in this study.

\subsection{Thermal Beaming from Rough Surfaces}
Real planets are not perfect isotropically-scattering spheres nor well-represented by a single blackbody emitter.  The thermal emission phase curves of airless rocky bodies in the solar system have a measured disk-integrated emission flux higher than expected from a perfect sphere at low phase angles (near eclipse for exoplanet observations), and a lower flux at large phase angles (e.g., \citealt{lebofsky86,hapke96,emery98,wohlfarth23}).  This effect, known as \textit{thermal beaming}, happens because hotter, more illuminated facets are preferentially tilted towards the host star at opposition, leading to thermal limb brightening.  This effect requires macroscopic roughness (typically at the scale of centimeters to millimeters) and negligible heat redistribution (extremely thin or no atmosphere).

\begin{figure*}
    \centering
    \includegraphics[width=1.0\linewidth]{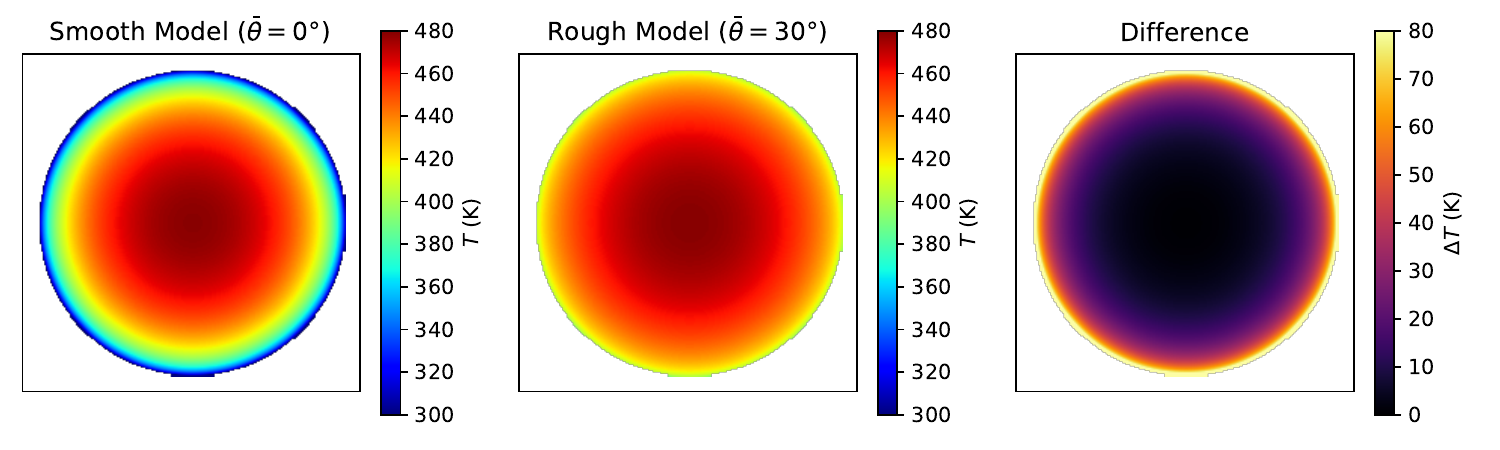}
    \caption{Effects of Moon-like macroscopic surface roughness on disk-resolved brightness temperatures at opposition (i.e., `thermal beaming') for TRAPPIST-1~c, assuming a root-mean-square roughness slope of $\bar{\theta}=30\degree$ (see \citealt{wohlfarth23} for details). Thermal beaming imparts a limb brightening-type effect that increases the disk-integrated brightness temperature (and thus eclipse depth) and may be detectable through phase curve observations \citep{zieba_lhsprop,tenthoff24}, but is largely indistinguishable from other thermal brightening/surface darkening effects for individual planets with current data.}
    \label{fig:roughmodel}
\end{figure*}

We use an advanced thermal roughness model from \citet{wohlfarth23} to quantify the effects of thermal beaming on exoplanet eclipse measurements.  Following \citet{wohlfarth23}, we assume a wavelength-albedo profile based on lunar basalt (synthesis of Chandryaan-1 Moon Mineralogy Mapper / M3 data and returned samples). The effects of surface roughness on disk-resolved brightness temperatures for TRAPPIST-1 c is shown in Fig. \ref{fig:roughmodel}. We find that, assuming Moon-like macroscopic roughness, this effect leads to roughly $5-15\%$ deeper eclipse depths in the LRS bandpass (dependent on the exact stellar and planetary emission spectra), corresponding to a $\sim2-5\%$ increase in brightness temperature. This is one possible explanation for hot ($T_{irr}\gtrsim\SI{1500}{K}$) planets where $\mathcal{R}\gtrsim1$---areas near the substellar point are expected to be partially or fully molten and have low albedo (e.g., \citealt{essack20}), while the colder limbs (where brightness temperature differences due to thermal beaming are the largest) remain solid, increasing the disk-integrated brightness temperature higher than that expected of a blackbody. 

Using secondary eclipse data alone, the effects of thermal beaming are difficult to distinguish from those of space weathering, as both lead to hotter brightness temperatures. In addition, the dependence of roughness on irradiation temperature is not obvious; sintering at high temperatures near the melting point of rock will likely smoothen surfaces, but closer-in worlds are likely subject to more frequent micrometeorite impacts that promote roughness. Thermal beaming may be directly detectable with spectroscopic phase curves \citep{zieba_lhsprop,tenthoff24}.

\subsection{Tidal Heating}

Tidal heating, due to tidal dissipation associated with both orbital eccentricity and rotational despinning, has been invoked as a theoretical pathway for dayside temperatures significantly hotter than that expected of a blackbody (e.g., \citealt{Crossfield22,lyu2024}). M-Earths in particular may be subject to high levels of tidal heating due to their close-in orbits \citep{driscoll2015}.

\subsubsection{Tidal Dissipation from Orbital Eccentricity}

JWST transit, eclipse, and phase curve observations allow for much more precise constraints on the eccentricity of rocky planets than previous transit timing variation or radial velocity measurements (e.g., \citealt{mahajan2024}).  All planets included in this study exhibit extremely low eccentricity values that are mostly $2\,\sigma$ within zero, consistent with a perfectly circular orbit given the Lucy-Sweeney bias on eccentricity \citep{lucy1971}. In addition, phase curve observations of M-Earths have all been consistent with zero nightside emission \citep{kreidberg19,zhang24,luque24}, whereas significant tidal heating is expected to produce detectable nightside emission (e.g., \citealt{lyu2024}).

Here, we use a so-called `fixed $Q$' model to estimate upper limits on the amount of tidal dissipation and the subsequent effects on $\mathcal{R}$ using reported eccentricity values. Following \citet{driscoll2015}, the tidal heat flux ($F_{tidal}$, in \si{W.m^{-2}}) is calculated via,
\begin{equation}
    F_{tidal}\times 4\pi R_{p}^{2}=-\frac{21}{2}\textrm{Im}(k_{2})G^{3/2}M_{\star}^{5/2}R_{p}^{5}\frac{e^{2}}{a^{15/2}},
\end{equation}
where $G$ is the gravitational constant, $M_{\star}$ is the stellar mass, $e$ is the orbital eccentricity.  $\textrm{Im}(k_{2})$ is the imaginary component of the second-order Love number which we set to mimic modern-day Earth [$\textrm{Im}(k_{2})=-0.003$].  This is equivalent to a tidal quality factor of $Q=100$ and $k_{2}$ value of 0.3. A fixed $Q$ model can be seen as a conservative estimate on the upper limit of heat flux via tidal dissipation, as this process becomes much less efficient as the mantle heats up to produce higher melt fractions \citep{driscoll2015}.

We note that there is large uncertainty in the tidal dissipation efficiency of close-in M-Earths. However, given the approximation $-\textrm{Im}(k_{2})\approx k_{2}/Q$, the efficiency of tidal dissipation assumed in our calculations is stronger than Moon-like ($k_{2}=0.024$, $Q=37.5$) and Mars-like ($k_{2}=0.164$, $Q=99.5$) bodies \citep{lainey2016}. Dynamical modeling of the TRAPPIST-1 system suggests that this ratio ($k_{2}/Q$) is less than or similar to Earth for the inner planets \citep{brasser2022}. In addition, interior structure modeling of the TRAPPIST-1 planets suggests that tidal heating, even at larger assumed eccentricities, has a negligible effect on their total energy budgets \citep{dobos19}.

\begin{figure}
    \centering
    \includegraphics[width=1.0\linewidth]{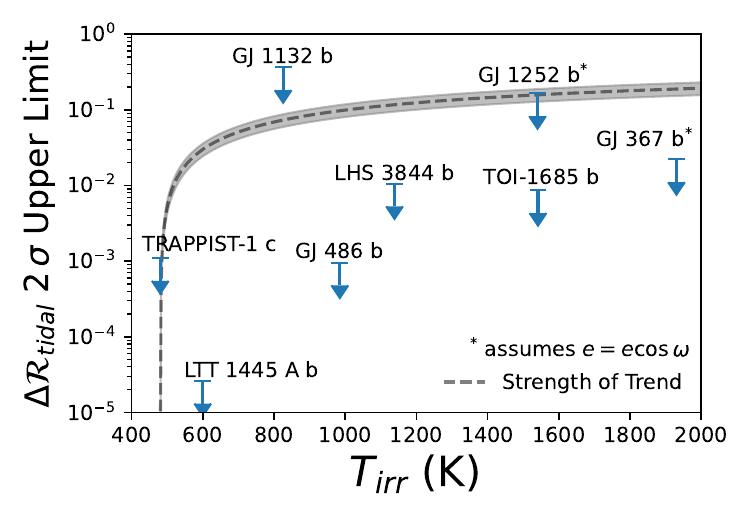}
    \caption{$2\,\sigma$ upper limits on the effects of tidal heating on $\mathcal{R}$ assuming Earth-like tidal dissipation ($Q=100,k_{2}=0.3$). We include the relative strength and $2\,\sigma$ uncertainties of the log-linear trend (using SPHINX models) explored in Section \ref{sec:results}, normalized to TRAPPIST-1~c. We find that the proposed trend likely cannot be directly explained by tidal dissipation due to orbital eccentricity.}
    \label{fig:tidal}
\end{figure}

Using each planet's $2\,\sigma$ upper limit on eccentricity, we then compare the tidal dissipation to the disk-averaged insolation received by the star assuming zero albedo ($F_{insol}$) to calculate the associated change in $\mathcal{R}$,
\begin{equation}
    \Delta\mathcal{R}_{tidal}=\left(\frac{F_{tidal}+F_{insol}}{F_{insol}}\right)^{1/4}.
\end{equation}
We exclude TRAPPIST-1 b as it does not have a reported eccentricity value from eclipse observations, although it is consistent with 0 \citep{greene23}.  We also calculate the circularization and tidal-locking (or synchronization) timescales $\tau_{circ}$ and $\tau_{sync}$ following \citet{gladman1996,jackson2008tidalevo}, where
\begin{equation}
    \tau_{circ}=\left(\frac{63}{4}\left(GM_{\star}^{3}\right)^{1/2}\frac{R_{p}^{5}}{Q_{p}M_{p}}\right)^{-1}a^{13/2},
\end{equation}
and
\begin{equation}
    \tau_{sync}=\frac{\omega_{p} a^{6}I_{p}Q_{p}}{3GM_{\star}^{2}k_{2}R_{p}^{5}},
\end{equation}
where $Q_{p}$ is the planet tidal quality factor (assumed to be 100), $M_{p}$ is the planet mass, $\omega_{p}$ is the initial planet rotation rate, and $I_{p}$ is the planet moment of intertia.  We set $\omega_{p}$ equal to $2\pi/P$ where $P$ is the orbital period, $I_{p}=0.4M_{p}R_{p}^{2}$, and $k_{2}=0.3$.

From these simple calculations, we show in Fig. \ref{fig:tidal} and Table \ref{tab:tidal} that tidal heating via orbital eccentricity alone is unlikely to directly explain the observed trend in $\mathcal{R}$ assuming an Earth-like interior structure.  Higher upper limits are associated with larger uncertainties in eccentricity and not necessarily higher tidal heat flux. In addition, orbital circularization timescales for planets in this study are short ($\tau_{circ}<10^{7}$ yr), with even shorter synchronization times ($\tau_{sync}<10^{3}$ yr) implying that eccentricities should be efficiently damped and that tidal locking should be extremely fast. Although, as discussed earlier, even small amounts of tidal heating may excite volcanic resurfacing, affecting the observed surface and potential atmospheric compositions.

\begin{deluxetable*}{lcccccc}
\tablecaption{Parameters and Results of a Fixed $Q$ Tidal Dissipation Model}
\tablewidth{0pt}
\tablehead{
\colhead{Planet} & \colhead{$a$} & \colhead{$M_{\star}$} & $\tau_{circ}$ & $\tau_{sync}$ & \colhead{Eccentricity} & \colhead{$\Delta\mathcal{R}_{tidal}$} \\[-2mm]
\colhead{} & \colhead{(AU)} &  \colhead{($M_{\odot}$)} & \colhead{(Myr)} & \colhead{(yr)} & \colhead{} & \colhead{Upper Limit}}
\startdata
TRAPPIST-1 c & 0.0158 & 0.09 & 1.3 &  80 & $0.0016^{+0.0015}_{-0.0008}$ & $1.1\times10^{-3}$ \\
TRAPPIST-1 b & 0.01154 & 0.09 & 0.16 &  19 & - & - \\
LTT 1445 A b & 0.0381 & 0.26 & 6.2 & 970 & $<0.0059$ & $2.6\times10^{-5}$\\
GJ 1132 b & 0.0157 & 0.1945 & 0.37 & 27 & $0.0118^{+0.0470}_{-0.0099}$ & $3.6\times10^{-1}$\\
GJ 486 b & 0.01714 & 0.312 & 0.32 & 23 & $0.00086^{+0.00160}_{-0.00043}$ & $9.5\times10^{-4}$\\
LHS 3844 b & 0.00622 & 0.15 & 0.0010 & 0.56 & $<0.001$ & $1.1\times10^{-2}$ \\
GJ 1252 b & 0.00915 & 0.38 & 0.0029 & 0.62 & $0.0025^{+0.0049}_{-0.0018}$$^{a}$ & $1.7\times10^{-1}$ \\
TOI-1685 b & 0.01138 & 0.454 & 0.010 & 1.9 & $0.0011^{+0.0013}_{-0.0007}$ & $8.6\times10^{-3}$\\
GJ 367 b & 0.00709 & 0.46 & 0.0028 & 0.35 & $0.0027^{+0.0008}_{-0.0008}$$^{a}$ & $2.2\times10^{-2}$ \\
\enddata
\tablecomments{Eccentricity values are from TRAPPIST-1 c \citep{zieba23}, LTT 1445 A b \citep{Wachiraphan24}, GJ 1132 b \citep{xue24}, GJ 486 b \citep{mansfield24}, LHS 3844 b \citep{lyu2024}, GJ 1252 b \citep{Crossfield22}, TOI-1685 b \citep{luque24}, and GJ 367 b \citep{zhang24}.  $^{a}$This is the reported $e\cos{\omega}$ value and thus may underestimate the true eccentricity.}\label{tab:tidal}
\end{deluxetable*}

\subsubsection{Asynchronous Rotation}
The above calculations assume perfect tidal locking, i.e. rotation synchronous with orbital period. Warm ($T_{irr}>\SI{400}{K}$) M-Earths are generally assumed to be tidally locked, due to extremely short tidal locking timescales from their close-in orbits (see Table \ref{tab:tidal}). While thick atmospheres may slow down the tidal locking process, modeling suggests that these planets are too close-in for this effect to be significant \citep{leconte2015}.  However, recent studies have also suggested that \textit{perfect} tidal locking is difficult (e.g., \citealt{leconte2018,revol2024}). Asynchronous rotation would result in a heating dependent on the rotation rate and heat capacity of the surface material (e.g., \citealt{lyu2024}), and as direct measurements of M-Earth rotation rates may be out of reach with JWST, this remains a source of uncertainty for the total amount of tidal heating. Future phase curve observations may be able to place more stringent constraints on the amount of tidal heating experienced by airless M-Earths. %via nightside emission.

%Although tidal locking for a rocky exoplanet has only been explicitly suggested for LHS~3844~b \citep{lyu2024}, follow-up phase curve observations may be able to confirm whether these planets are tidally locked.

\subsection{Hazes}\label{sec:aerosols}

The atmospheres tested in the thin atmosphere hypothesis of Section \ref{sec:atms} were assumed to be clear (absent of aerosols or clouds).  However, aerosols and clouds are ubiquitous in the atmospheres of rocky planets in the solar system (including Titan) and have been observed to be also prevalent in (albeit non-``rocky'') exoplanets that span similar equilibrium temperatures to those analyzed here \citep{kempton23, beatty24}.

Hazes in a planet's dayside upper atmosphere can efficiently absorb incident radiation at shorter wavelengths, creating a thermal inversion---i.e., a stratosphere that is hotter than at pressures deeper below (e.g., \citealt{hu12_photochemistry}).  This ``flips'' atmospheric features to be observed in emission rather than absorption, as is seen on Titan (e.g., \citealt{coy2023}).  For rocky planets around M stars in particular, even the near infrared molecular absorption of H$_2$O can cause said inversions \citep{malik19rocky}, but hazes can potentially be even more efficient absorbers and have a greater impact on the thermal profile.  Given that such hazes form via UV photochemistry on Titan (e.g., \citealt{nixon2024}), it is possible that they also readily form around M-Earths that receive high amounts of UV radiation \citep{peacock19}.

For narrow band observations covering specific absorption features (i.e., MIRI F1500W), the inverted emission feature could potentially cancel out the effects of redistributive cooling and lead to a low inferred albedo. This mechanism has been invoked to explain blackbody-like $\mathcal{R}$ values for MIRI F1500W observations despite potentially having a \ce{CO2}-rich atmosphere \citep{ducrot23}.  Titan's haze is formed through complex processes that begin with photolysis of methane (\ce{CH4}) and nitrogen gas (\ce{N2}).  Laboratory experiments have shown that hazes can form in \ce{CO2}-dominated, hydrocarbon-poor environments \citep{horst2018,he20}. However, these experiments have also suggested that haze formation efficiency in such environments is much less than Titan-like conditions and thus the feasibility of extremely hazy \ce{CO2}-dominated atmospheres remains unclear.

\begin{figure}
    \centering
    \includegraphics[width=1.0\linewidth]{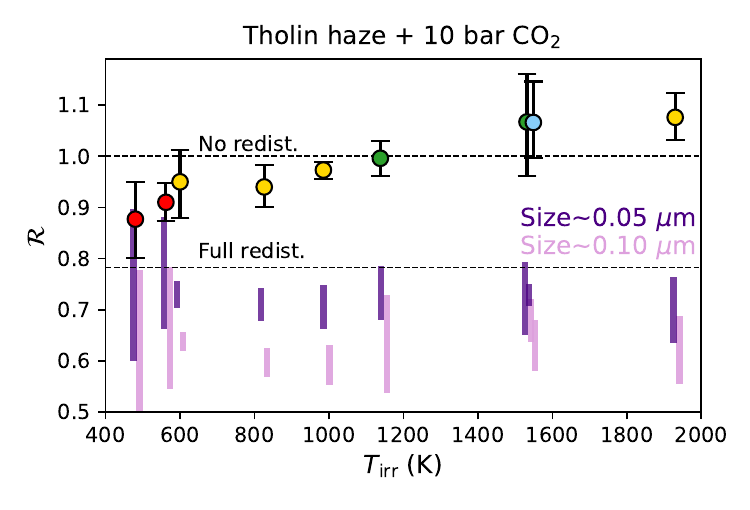}
    \caption{Modeled brightness temperature ratios ($\mathcal{R}$) per planet-instrument setup of thick, hazy atmospheres, for two particle size distributions of Titan-like tholins.  The range in $\mathcal{R}$ for each planet spans the modeled volume mixing ratios of haze particles ($10^{-12}-10^{-6}$), assumed to be vertically constant.  The rest of the atmosphere is comprised of 10 bars of CO$_2$.  Brightness temperatures are calculated in the bandpass of each instrument, with TRAPPIST-1~b accounting for both MIRI F1280W and F1500W. While Titan-like hazes can lead to low-inferred albedo for narrowband photometric observations (i.e., MIRI F1500W), we do not find a false negative scenario for thick atmospheres for any other planet-instrument pair.
    }
    \label{fig:haze}
\end{figure}

\begin{figure*}
    \centering
    \includegraphics[width=1.0\linewidth]{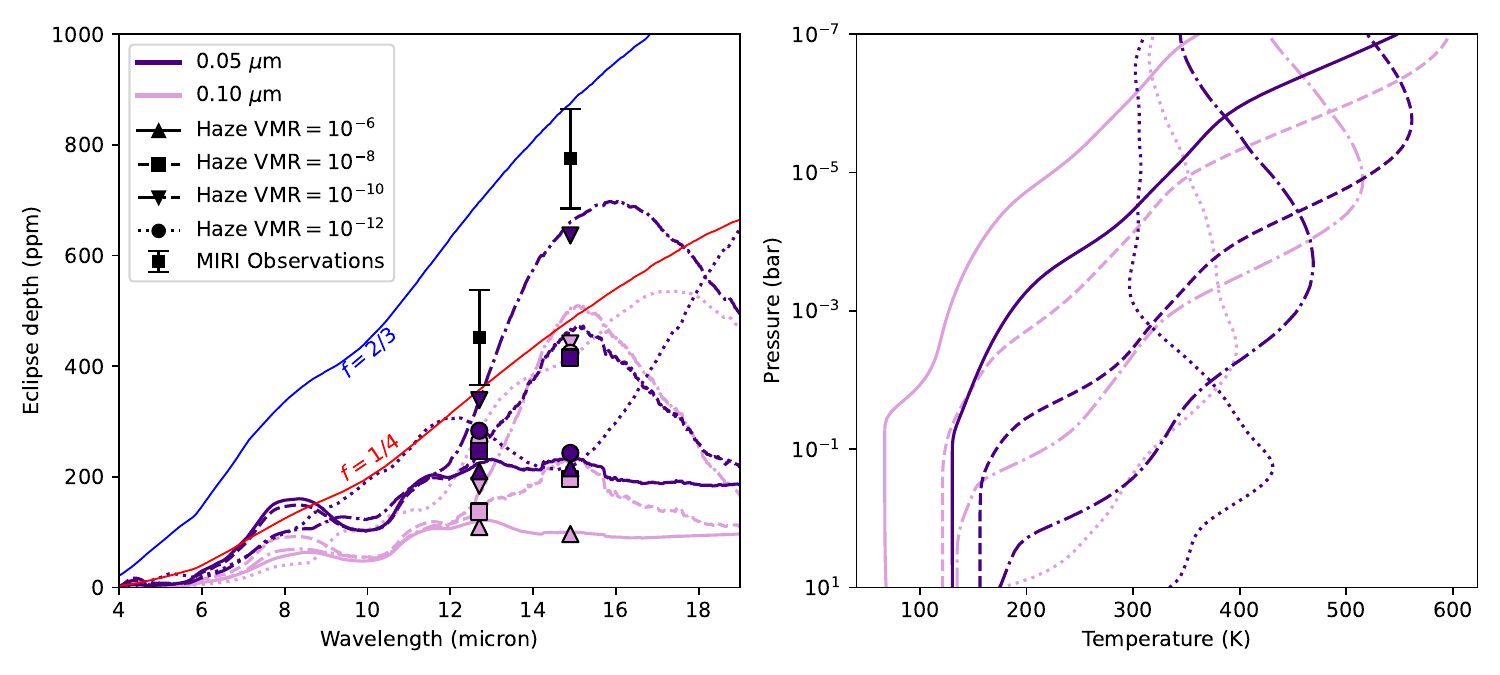}
    \caption{Example \texttt{HELIOS} emission spectra and associated temperature-pressure profiles for TRAPPIST-1 b, assuming a thick 10-bar \ce{CO2} atmosphere with Titan-like tholin haze volume mixing ratios of $10^{-12}-10^{-6}$, compared to blackbody models assuming no ($f=2/3$) and full ($f=1/4$) heat redistribution to the nightside.  While several of these atmospheres show strong thermal inversions, we find that only one tested scenario (VMR of $10^{-6}$, particle size mean of $0.05\,\micron$) exhibits a false negative for a thick atmosphere.}
    \label{fig:hazet1b}
\end{figure*}

To test whether this effect can produce a false negative for a thick atmosphere, we forward model 10 bar CO$_2$ atmospheres that have Titan-like tholin hazes, using the optical properties from the OptEC$_{\textrm{(s)}}$ model \citep{jones13_advanced} used in \citet{ducrot23}. We assume a fiducial value for the optical band gap of 2.0 eV \citep{imanaka04_laboratory}.  In lieu of detailed photochemical and microphysics modeling, we focus on end member scenarios and assume vertically fixed volume mixing ratios (VMRs) of $10^{-6}-10^{-12}$ in 2-dex intervals.  This roughly encompasses the range of haze VMR observed in Titan's mid-to-lower atmosphere \citep{fan19}.  We use two log-normal particle size distributions of varying means of 50 nm and 100 nm to match lab experiment results in \citet{he20}, with a geometric standard deviation of 1.1.  We show the range of derived $\mathcal{R}$ per planet-instrument pair in Fig. \ref{fig:haze}.  For the photometric observations of TRAPPIST-1 planets, we find that a low inferred albedo could be explained by a haze-induced thermal inversion (see Fig. \ref{fig:hazet1b}).  However, we find that this effect is of less concern for observations of other planets due to instrument choice and irradiation temperature.  For instruments that capture a significant amount of the planet's total thermal emission like MIRI LRS, the continuum typically probes into colder regions of the atmosphere (Figs. \ref{fig:haze} \& \ref{fig:hazet1b}, also \citealt{malik19helios}).  Secondly, for planets with hotter irradiation temperature, there is less spectral separation between the incident stellar flux and the outgoing thermal emission; as the deposited energy in the atmosphere is reprocessed more uniformly through the atmosphere, the thermal gradient is reduced \citep{guillot10}.

We note that the effect of hazes on the atmospheric thermal structure is highly dependent on model assumptions such as the VMR, particle size distribution, and band gap, as most of our tested scenarios \textit{do not} exhibit low inferred albedo. In addition, hazes can introduce high Bond albedo (e.g., \citealt{kempton23}) that would further cool the dayside effective temperature.

We also simulated observations using the Titan-like tholin haze models from \citet{kitzmann18}, which have similar extinction cross sections as the OptEC$_{\textrm{(s)}}$ model below 2 micron but have roughly one order of magnitude larger absorption cross sections past 2 micron, i.e., less contrast between the short and longwave absorption.  We find that with this haze model we are unable to produce thermal inversions strong enough to match the observations of the TRAPPIST-1 planets.

To summarize, we find that false negatives for thick atmospheres due to haze-induced thermal inversions are unlikely, and in the case that they are present, broadband observations are useful to distinguish between a thermal inversion and weak heat redistribution scenario. JWST Cycle 2 phase curve observations of TRAPPIST-1 b and TRAPPIST-1 c (GO-3077, PIs: Gillon \& Ducrot) will be able to definitively determine whether they host such extremely hazy atmospheres.

\subsection{Nightside Clouds}
Another effect that can potentially affect the dayside heat budget and give rise to a false negative is the preferential formation of clouds on the nightside of the planet.  Such clouds, if optically thick, can inhibit radiative cooling and trap the heat via efficient greenhouse effect on the nightside, resulting in a net warming \citep[e.g.,][]{turbet21}.  Additionally, the clouds give rise to the nightside emission emerging from lower pressures than the dayside emission; this could in turn exaggerate the day/night brightness temperature contrast relative to the actual temperatures at the surface  \citep{powell24}. However, a false negative detection via eclipses alone (i.e., a near blackbody-like $\mathcal{R}$) due to this effect requires an absence of IR-absorbing gases (e.g., \ce{CO2}, \ce{H2O}, \ce{CH4}) thought to be common in terrestrial atmospheres.  More work is required to establish in what regimes nightside clouds can be false negatives for thick atmospheres.

\section{Discussion} \label{sec:discussion}

\subsection{Cosmic Shoreline Hypothesis}\label{sec:shoreline}

The Cosmic Shoreline hypothesis posits that whether solar system bodies are able to retain significant atmospheres or not is controlled by atmospheric escape processes, not initial volatile endowment: ``nurture'', not ``nature'' \citep{zahnle17}. The Cosmic Shoreline has been widely invoked in observations of M-Earths, (e.g., \citealt{may23,moran23,xue24,mansfield24,Redfield2024,Wachiraphan24}), as it presents an empirically motivated population-level prediction of escape theory that can be formally tested with observations. This formalism also underscores much of the current framework of the Rocky Worlds DDT program. It is still unclear which of XUV radiation, bolometric instellation (radiation over all wavelengths), or high-energy impactors is the main control on atmospheric loss for Earth-sized planets \citep{schlichting18, wyatt20, king21}.

M-Earths endure very high XUV flux, especially in the pre-saturation phase of their host stars, though stellar activity that can drive atmospheric loss continues for many gigayears into the main sequence \citep{king21}. In addition, their close-in orbits subject them to high-energy impactors, further promoting atmospheric loss \citep{wyatt20}.  However, late-stage impactors, especially for colder M-Earths, might replenish atmospheres depreciated by XUV and high-energy impacts \citep{kral18}.

\begin{figure}
    \centering
    \includegraphics[width=1\linewidth]{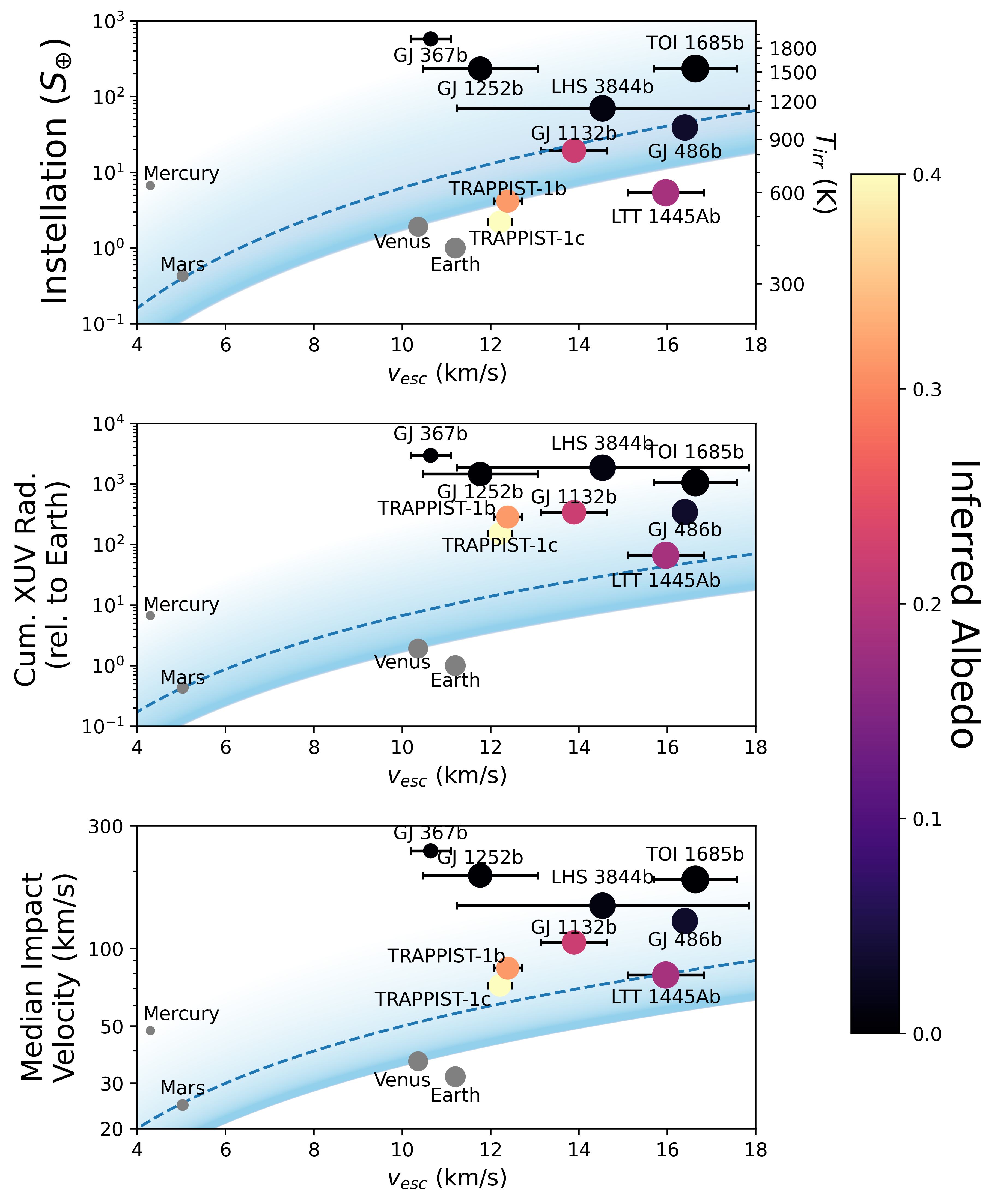}
    \caption{Observed inferred albedo values in the context of the Cosmic Shoreline \citep{zahnle17} hypothesis, in (top) instellation, (middle) estimated cumulative XUV radiation based on methods outlined in Appendix \ref{sec:uvmodel}, and (bottom) median impact velocity space. Following \citet{zahnle17}, median impact velocity is estimated as $v_{imp}\approx \sqrt{v_{esc}^{2}+v_{orb}^2}$, where $v_{orb}$ is the Keplerian orbital velocity.  The Shorelines follow the scaling laws (top) $v_{esc}\propto I^{4}$, (middle) $v_{esc}\propto I_{\mathrm{XUV}}^{4}$, and (bottom) $v_{esc}=5\,v_{imp}$.  The dotted line is normalized to Mars, whereas the shaded regions are calibrated to Venus and Mercury.  The is large uncertainty in the exact shoreline position; it is still unclear where the shoreline should lie for M-Earths, and whether it is narrow or wide.
    }
    \label{fig:shoreline}
\end{figure}

In Fig. \ref{fig:shoreline}, we show our sample of planets in the context of these three loss mechanisms alongside solar system planets. These results suggest that these planets are not expected to be able to retain significant atmospheres from multiple atmospheric loss standpoints, consistent with their measured near blackbody-like $\mathcal{R}$ values.

However, the solar system is unusual [e.g., most exoplanetary systems lack a Jupiter analog \citep{fernandes19}, and the occurrence rate is even lower for the M-type stars considered in this study \citep{montet14}], so the reliance of these three scaled approaches on solar system data introduces major uncertainty.  As \citet{zahnle17} wrote, 
``we do not know if the shoreline is broad or narrow (i.e., whether the transition from a thin atmosphere to an atmosphere too thick and deep to be habitable to an ecology like our own is gentle or abrupt), nor in which ways our solar system is representative or unrepresentative of extrasolar systems.'' Fig. \ref{fig:shoreline} also does not consider processes that can replenish volatiles over long timescales [e.g., late-stage prolonged outgassing \citep{kite20} or cometary impacts \citep{kral18}]. Thus, more data are needed to test if the tendency for planets closer to the Shoreline to generally show higher inferred albedo corresponds to a Cosmic Shoreline, or something else.

\subsection{Can Reflective Bare-Rock Surfaces be False Positives for Atmospheres?}\label{sec:false_pos}

\citet{mansfield19} argued that for warm ($\SI{400}{K}<T_{irr}<\SI{1250}{K}$) rocky planets, bare-rock surfaces (originally presented in \citealt{hu12}) are unlikely to serve as false positives for atmosphere detection. However, the energy balance model of \citet{mansfield19} did not distinguish between geometric and spherical albedo, leading to estimated planetary Bond albedos that range from $\sim$70\% to $\sim100\%$ that of the true value.

\begin{figure*}
    \centering
    \includegraphics[width=0.66\linewidth]{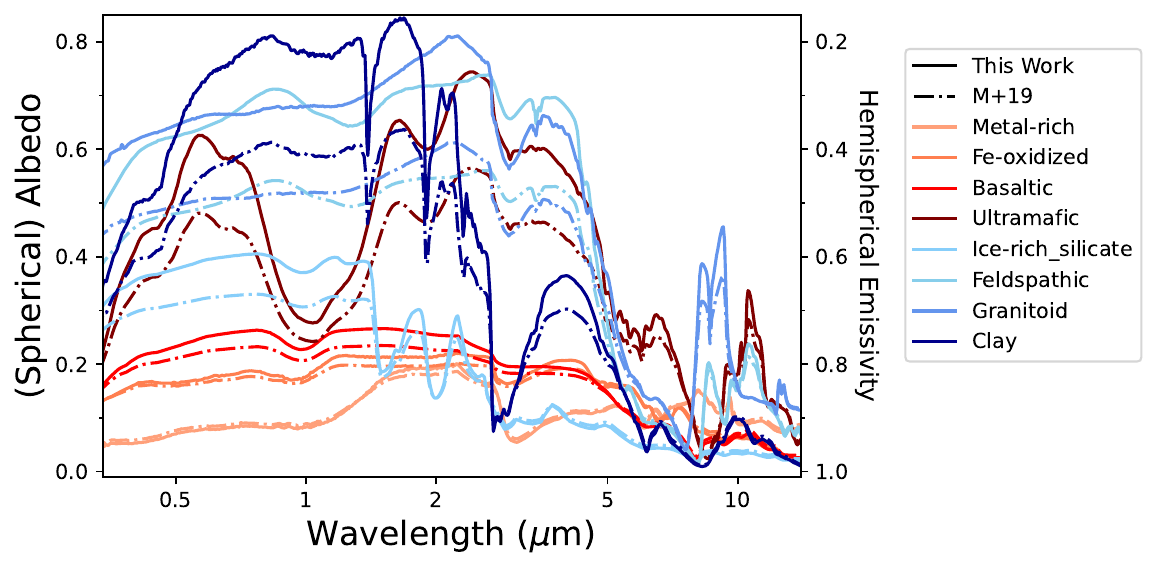}
    \caption{Comparison of spherical albedo values calculated using data from \citet{hu12} used in this study (solid lines) with those in \citet{mansfield19} (dash-dotted lines).  Colors represent surface types from \citet{hu12}.}
    \label{fig:m19_comparison}
\end{figure*}

\textit{Geometric} albedo refers to the fraction of light reflected (and emitted) towards the observer when compared to a perfect Lambertian disk.  This value is fundamentally different from the \textit{spherical} albedo, which describes the total fraction of incoming light scattered in all directions. The Bond albedo, and subsequently the amount of radiation absorbed by the planet, is determined by integrating the spherical albedo over the incoming stellar spectrum.  Hemispherical emissivity is also determined via the spherical albedo.  These quantities are all fundamentally computed from the single scattering albedo of the material, which we use in this work (Section \ref{sec:calc_albedo}).  An explicit comparison of albedo profiles used in this study and those used in \citet{mansfield19} is shown in Fig. \ref{fig:m19_comparison}.

The increased albedos used in our model cool the planet, leading to higher values of inferred albedo for most bare-rock surfaces than presented in \citet{mansfield19}.  This effect is also noticeable in the recent work of \citet{hammond24}, who derived relatively high ($>0.34$) Bond albedos for a wide variety of plausible surface types from the RELAB spectral database.

For some surface types, this cooling effect can be comparable to the heat redistribution expected of thick atmospheres.  For TRAPPIST-1, our derived Bond albedos for the surface profiles from \citet{hu12} are as follows: metal-rich - 0.13,  Fe-oxidized - 0.20,  basaltic - 0.24, ultramafic - 0.49, feldspathic - 0.67, clay - 0.67, and granitoid - 0.69.  Inferred albedos in the LRS bandpass are similarly high (Fig. \ref{fig:spaceweather}).  The inferred albedo of some of these surfaces is comparable to the cooling expected of full heat redistribution for a thick atmosphere ($A_{i}=0.63, \mathcal{R}=0.78$), suggesting that some plausible geological surfaces (in the absence of space weathering) may serve as false positives for atmospheric heat redistribution.  This was also recently highlighted by \citet{hammond24}.

However, the effect of surface composition on the Bond albedo of airless planets is likely overestimated by the albedo profiles of \citet{hu12}. Mercury and the Moon have low Bond albedos of 0.06 and 0.13, respectively, despite having fine-grained, largely basaltic regolith \citep{mallama2002,matthews2008}.
The discrepancy between theoretical and observed albedos is due to space weathering on the outermost layer of the surface. Indeed, we find that the Bond albedo of an example pulverized lunar basalt (RELAB ID: LR-CMP-158, originally analyzed in \citealt{pieters16}) for a Sun-like star is 0.23, about twice that of lunar surface soils.  Darkening from space weathering is ubiquitous for solar system airless bodies: $\sim85\%$ of near-Earth asteroids exhibit a visual geometric albedo of less than 0.3 despite being petrologically diverse \citep{wright16,morbidelli20}.  These values are overestimated compared to the true Bond albedo due to the opposition surge effect  \citep{belskaya00}. Bond albedos are typically $\sim40\%$ that of visual geometric albedo given standard assumptions for asteroids \citep{muller12}, implying $85\%$ of asteroids have a Bond albedo $\lesssim0.12$.

In addition, when considering the effects of fine-grained regolith on the Bond albedo of airless M-Earths, one must also consider the effects of space weathering on their surfaces. As stated before, regolith formation on planet-sized airless bodies (such as the Moon and Mercury) is thought to be governed by meteorite/micrometeorite impacts and solar wind bombardment \citep{melosh1989,mckay1991,domingue2014}.  This suggests that space weathering, which is also caused by micrometeorite impacts and solar wind bombardment \citep{hapke01}, occurs via the same pathways as regolith formation.  The high Bond albedos of surfaces presented in \citet{hu12} and \citet{hammond24} require fine-grained regolith, as larger-grained or natural slab surfaces show much lower albedo for the same underlying material (e.g., \citealt{zhuang23,paragas25}).  However, these studies do not consider the potential darkening effects of space weathering and their impact on brightness temperatures. Future studies of the geological plausibility of the surface types presented in \citet{hu12} and the expected degree of space weathering on airless M-Earths will be needed to more carefully assess the risk of false positive atmospheric detections through secondary eclipse measurements.

\subsection{Future Tests}

\begin{figure*}
    \centering
    \includegraphics[width=0.6\linewidth]{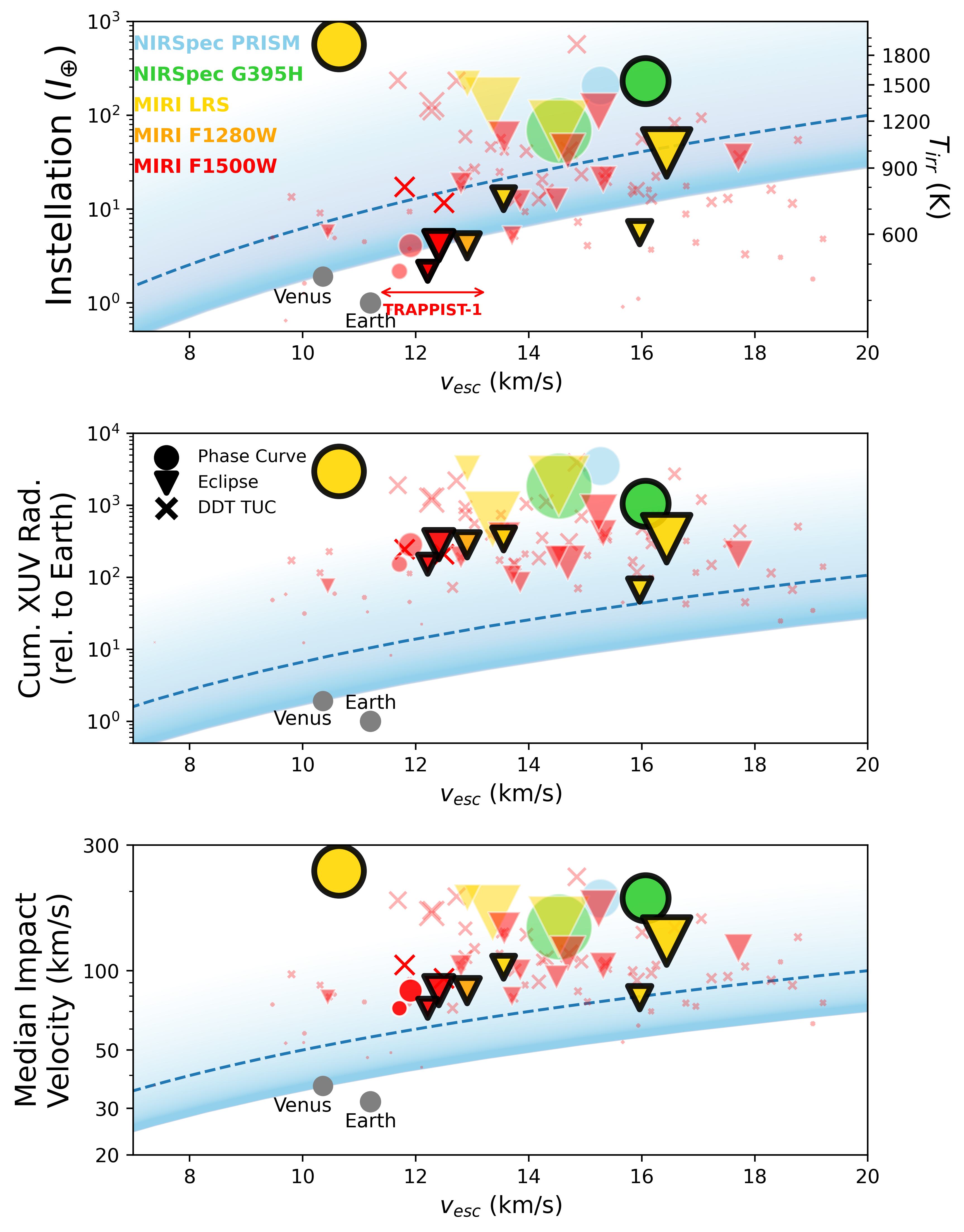}
    \caption{Scheduled M-Earth emission observations for JWST Cycle $1-3$ and Rocky Worlds DDT targets under consideration (TUCs) in the context of the Cosmic Shoreline hypothesis: in (top) instellation, (middle) estimated cumulative XUV radiation (see Appendix \ref{sec:uvmodel}), and (bottom) median impactor velocity space. As in Fig. \ref{fig:shoreline}, the dotted line is normalized to Mars and the shaded regions are calibrated to Venus and Mercury. Point area represents the estimated relative SNR per instrument+planet setup; SNR is calculated as $\mathrm{ESM}\times\sqrt{N_{obs}}$, where $N_{obs}=N_{eclipse}+4N_{phase curve}$ (10 eclipses were assumed for TUCs). Planets with black borders represent observations included in this study, all of which lack compelling evidence for thick atmospheres. Separate TRAPPIST-1 observations are offset horizontally for clarity. All data for DDT targets and Cycle $1-3$ observations not included in this study, except estimated system ages, are from the NASA Exoplanet Archive. Despite many planets seeming like promising candidates for atmospheric retention in terms of their instellation, the high cumulative XUV of their low-mass host stars and high susceptibility to impact-based atmospheric erosion leads to a more pessimistic outlook for the M-Earth opportunity.  }
    \label{fig:cycle123}
\end{figure*}

JWST Cycles $1-3$ include emission observations of $\sim25$ M-Earths.  Most of these observations aim to detect or rule out the presence of an atmosphere, either through direct detection of \ce{CO2} spectral features, or indirect detection via atmospheric heat transport.  Fig. \ref{fig:cycle123} shows that several of these targets appear favorable when plotted against the instellation-based `Cosmic Shoreline'.  However, almost all M-Earth targets experienced much more XUV radiation than Earth (due to their low-mass host stars) and likely experienced frequent high-energy impacts (due to their close-in orbits): both are detrimental for atmospheric retention.  These results are timely for informing the upcoming 500 hour Rocky Worlds DDT program to search for atmospheres on M-Earths with MIRI F1500W \citep{Redfield2024}.  

Thus, if the Rocky Worlds DDT fails to find evidence for a substantial M-Earth atmosphere on any target [i.e., a shallow eclipse depth unlikely to be caused by surface mineralogy as defined in \citet{mansfield19} or systematic effects], this would suggest that either cumulative XUV radiation or the energy of impactors, both highly elevated for M-Earths compared to terrestrial planets around FGK stars, are effective at removing atmospheres of small planets.  Another possibility is that M-Earths form volatile poor compared to rocky planets in the solar system (e.g., \citealt{desch2020,mulders2015}). Such results would motivate atmosphere searches for rocky planets around higher-mass stars. However, emission observations of airless M-Earths would remain useful in probing geological surface processes and crustal compositional diversity of rocky exoplanets, as well as constraining their outgassing and atmospheric loss histories (e.g., \citealt{foley2024,first2024,paragas25}).  As shown in this work, this can be done by assessing trends in brightness temperature.

\section{Conclusions} \label{sec:summary}
Secondary eclipse data for M-Earths show a tentative trend in the brightness temperature ratio $\mathcal{R}$ (the dayside brightness temperature ratioed to that of a perfect blackbody) as a function of irradiation temperature, with hotter planets exhibiting lower inferred albedo. The trend is strongly favored statistically over no trend when using the most recent stellar models available for M dwarfs. However, the statistical evidence is dependent on the stellar model and method used to derive brightness temperature, and we consider its identification tentative. Options to explain this trend include: 
\begin{itemize}
  \item Larger regolith grain sizes caused by higher rates of volcanic resurfacing on close-in, hotter planets or grains sintering at high temperatures;  
  \item Space weathering via micrometeorite impactors and stellar winds darkening faster-orbiting, closer-in planets.  However, the degree of space weathering for outer planets would have to be much less than that predicted by the stellar wind strength scaling in \citet{zieba23};
  \item Colder, outer planets retain thin ($<1$ bar) \ce{CO2}-rich outgassed atmospheres while closer-in planets lose (or never gain) such atmospheres quickly due to high atmospheric loss fluxes. However, such atmospheres would likely be subject to atmospheric collapse on the nightside and require constant resupply via volcanism.
\end{itemize}

We show that, assuming an Earth-like interior, tidal dissipation due to orbital eccentricity is unlikely to directly explain the proposed trend due to the extremely low eccentricities of planets in this study.
We also show that Titan-like aerosols in \ce{CO2}-dominated atmospheres have potential, but are unlikely, to serve as false negatives for atmospheres through photometric eclipse observations.

Future observations with JWST will begin to fill in unexplored parameter space (see Fig. \ref{fig:cycle123}), which will help to break the degeneracies between each scenario through multi-band or spectroscopic characterization. If the trend is \textit{geological}, these observations may constrain surface properties such as the mineralogy and grain size through silicate absorption features like the Si-O stretch 
(e.g., \citealt{shirley2019,first2024,paragas25}).  Spectroscopic phase curves have the potential to directly constrain the degree of space weathering and roughness of regolith surfaces \citep{tenthoff24}. If the trend is \textit{atmospheric}, these observations may constrain the presence/absence of gaseous absorption features, while phase curves may directly probe the amount of heat redistribution from thick atmospheres. Leveraging the statistical advantages of much larger sample sizes, these observations will provide further constraints on the location of the M-Earth `Cosmic Shoreline'.

\section*{Acknowledgments}
%\begin{acknowledgements}
B.P.C. was supported through a NASA grant awarded to the Illinois/NASA Space Grant Consortium. J.I. acknowledges funding from the Alfred P. Sloan Foundation under grant G202114194. D.K. acknowledges support from the National Science Foundation of China (NSFC) grant 12473064. M.W.M. and M.Z. acknowledge support from the Heising-Simons Foundation through the 51 Pegasi b Fellowship Program.  E.M.-R.K. acknowledges support from the NASA Exoplanets Research Program under grant 80NSSC24K0157.

This research has made use of the NASA Exoplanet Archive, which is operated by the California Institute of Technology, under contract with NASA under the Exoplanet Exploration Program.  This research utilizes spectra acquired by John F. Mustard and Carle M. Pieters with the NASA RELAB facility at Brown University. The authors thank Mark Hammond for discussions on surface albedo profiles and Elsa Ducrot and Michiel Min for providing their OptEC(s) haze model. We thank an anonymous reviewer for comments that helped improve the manuscript.

%\end{acknowledgements}

\appendix
\restartappendixnumbering 
\section{Data Considerations for Individual Planets} \label{ap:data_considerations}

\subsection{TRAPPIST-1 b: F1500W vs. F1280W} \label{sec:ap_trappist}
Early eclipse observations of TRAPPIST-1 b with MIRI F1500W had originally suggested a blackbody-like brightness temperature ($\mathcal{R}_{15\mu\mathrm{m}}\approx 0.99\pm0.05$, \citealt{greene23}).  This, however, is at odds with more recent eclipse data with F1280W, which suggests a lower brightness temperature at \SI{12.8}{\micron}  ($\mathcal{R}_{12.8\mu\mathrm{m}}\approx 0.85\pm0.06$, \citealt{ducrot23}).  Furthermore, reanalysis of the F1500W data suggests a slightly lower eclipse depth, refining the 15 micron brightness temperature to $\mathcal{R}_{15\mu\mathrm{m}}\approx 0.95\pm0.05$ \citep{ducrot23}.  In this work, we use the most up-to-date reduction of both F1280W and F1500W data from \citet{ducrot23} (giving a combined $\mathcal{R}\approx 0.91\pm0.04$), although we include the original \citet{greene23} F1500W eclipse depth results in Fig. \ref{fig:trend} and Table \ref{tab:Rvalues}.  TRAPPIST-1 b is the target of future emission observations with JWST using F1500W [GO-3077 (PI: Gillon \& Ducrot), GO-5191 (PI:Ducrot)], which will help give more precise constraints on its dayside brightness temperature.

\subsection{TOI-1685 b: NRS1 vs. NRS2}
NIRSpec G395H phase curve observations of TOI-1685 b show strong levels of correlated noise \citep{luque24}. These signals seem to be detector (NRS1 vs. NRS2) dependent and systematic rather than astrophysical in origin.  To account for this, \citet{luque24} performed a prayer-bead analysis of the white-light phase curves and the eclipse emission spectra that increases uncertainty estimates to account for this correlated noise.

According to \citet{luque24}, the NRS2 detector is much less affected by this correlated noise and thus we use the NRS2 prayer-bead analysis white-light phase curve amplitude for our main analysis and the prayer-bead eclipse spectrum for spectral analysis.  We also adopt the reported NRS2 prayer-bead analysis' planet-to-star radius ratio ($R_{p}/R_{\star}$) during nested sampling, as it conflicts significantly with previous results from \citet{burt2024}.

\subsection{MIRI F1500W Systematics in Observations of LHS~1478~b}\label{sec:ap_lhs}
Recent results from the Hot Rocks Survey \citep{hotrocks} report a shallow MIRI F1500W eclipse depth for the warm ($T_{irr}=\SI{840}{K}$) M-Earth LHS~1478~b \citep{august2024}. The reported eclipse depth ($138\pm53$ ppm) is less than half that expected of a blackbody, and implies an $\mathcal{R}$ value of $0.66\pm0.10$, \textit{significantly} lower than any other planet considered in this study. However, these results are complicated by the presence of strong correlated noise in the second visit.  Without the use of Gaussian processes to remove this correlated noise, the second eclipse produces a negative eclipse depth that is more than $5\,\sigma$ inconsistent with the first.
With the use of Gaussian processes for noise removal, \citet{august2024} still failed to recover the second eclipse.  For these reasons, we do not include results from \citet{august2024} in this study. All other planets in this study have shown repeatable or very high confidence ($\gtrsim5\,\sigma$) eclipse detections.

\subsection{GJ 367 b Anomalous 9 Micron Emission Feature}
Following \citet{zhang24}, for spectral fitting we disregard the anomalously deep eclipse depth at $9\micron$, which they attribute to poorly-understood systematics. This does not affect the white-light results.

\begin{deluxetable*}{lcccccc}
\tablecaption{Comparison of $\mathcal{R}$ with Previous Studies}\label{tab:comparison}
\tablewidth{0pt}
\tablehead{
\colhead{Planet} & \colhead{$T_{irr}$} & \colhead{Eclipse Depth(s)} & \multicolumn{2}{c}{$\mathcal{R}$} & 
\colhead{Instrument(s)} & \colhead{$M_p$} & \colhead{$R_p$} \\[-2mm]
\colhead{} & \colhead{(K)} & \colhead{(ppm)} & \colhead{SPHINX} & \colhead{PHOENIX} & \colhead{} & \colhead{$(M_{\oplus})$} & \colhead{$(R_{\oplus})$}}

\tablehead{\colhead{Planet} & \colhead{Reference} & \colhead{Reported $\mathcal{R}$} & \colhead{Reduction Used} &
\multicolumn{2}{c}{$\mathcal{R}$ (This Work)} \\[-2mm]
\colhead{} & \colhead{} & \colhead{} & \colhead{(If Applicable)} & \colhead{SPHINX} & \colhead{PHOENIX}}
\startdata
TRAPPIST-1 c & \citet{zieba23} & $0.88\pm0.07^{a}$ & - & $\mathbf{0.877^{+0.073}_{-0.075}}$ & $0.903^{+0.075}_{-0.082}$ \\
TRAPPIST-1 b$^{b}$  & \citet{greene23} & $0.99\pm0.05^{a}$ & F1500W Only & $0.993^{+0.049}_{-0.052}$ & $1.021^{+0.054}_{-0.055}$ \\
LTT 1445 A b & \citet{Wachiraphan24} & $0.952\pm0.057$ & \texttt{SPARTA} &$\mathbf{0.950^{+0.063}_{-0.071}}$ & $0.955^{+0.066}_{-0.072}$\\
GJ 1132 b & \citet{xue24} & $0.95\pm0.04$ & \texttt{SPARTA} & $\mathbf{0.940^{+0.043}_{-0.040}}$ & $\mathbf{0.952^{+0.042}_{-0.044}}$ \\
GJ 486 b & \citet{mansfield24} & $0.97\pm0.01$ & \texttt{SPARTA} & $\mathbf{0.973^{+0.016}_{-0.017}}$ & $\mathbf{0.978^{+0.016}_{-0.015}}$ \\
LHS 3844 b & \citet{kreidberg19} & $1.01\pm0.05^{a}$ & - & $0.996^{+0.033}_{-0.034}$ & $\mathbf{1.002^{+0.033}_{-0.034}}$ \\
GJ 1252 b & \citet{Crossfield22} & $1.01^{+0.09}_{-0.11}$$^{a}$ & - & $1.067^{+0.094}_{-0.105}$ & $\mathbf{1.035^{+0.090}_{-0.103}}$ \\
TOI-1685 b & \citet{luque24} & $0.98\pm0.07$ & Prayer-bead NRS2 & $1.066^{+0.080}_{-0.069}$ & $\mathbf{1.008^{+0.076}_{-0.058}}$ \\
GJ 367 b$^{b}$ (Spectral) & \citet{zhang24} & $0.99\pm0.06^{a}$ & \texttt{SPARTA} & $1.002^{+0.049}_{-0.045}$ & $0.966^{+0.044}_{-0.039}$ \\
\enddata
\tablecomments{Bolded values indicate the stellar model originally used to derive reported $\mathcal{R}$ values. $^{a}$This value is derived from the reported brightness temperature (in Kelvin) and uncertainties. $^{b}$\citet{greene23} derived brightness temperature from the absolute flux independent of stellar model whereas \citet{zhang24} used the observed stellar spectrum. }
\end{deluxetable*}

\section{XUV Model}
\label{sec:uvmodel}

To estimate the cumulative XUV irradiation experienced by the planets in this work, we follow the simplified broken power law approach of \citet{rogers21}.  In this framework,

\begin{equation}
\frac{L_{\mathrm{XUV}}}{L_{\text{bol}}} = 
\begin{cases} 
10^{-3.5} \left(\frac{M_{*}}{\mathrm{M}_{\odot}}\right)^{-0.5} & \text{for } t < t_{\text{sat}}, \\
10^{-3.5} \left(\frac{M_{*}}{\mathrm{M}_{\odot}}\right)^{-0.5} \left(\frac{t}{t_{\text{sat}}}\right)^{-1.5} & \text{for } t \ge t_{\text{sat}}.
\end{cases}
\end{equation}
where $L_{bol}$ is the bolometric luminosity, $L_{XUV}$ is the XUV luminosity, $M_{\star}$ is the star mass, and $t_{sat}$ is the `saturation time' defined by
\begin{equation}
t_\text{sat} = 10^{2} \, \bigg (\frac{M_*}{\mathrm{ M}_\odot } \bigg)^{-1.0} \, \text{Myr}.
\end{equation}

We use empirical isochrones derived for low-mass stars from \citet{herczeg15} to estimate host stars' bolometric luminosity as a function of mass and time. We highlight that, as noted in \citet{zahnle17}, the pre-saturation phase is the dominant contributor to the total XUV radiation.  The model is then run through for the entire estimated system age, ignoring error estimates for simplicity (see Table \ref{tab:ages}).  Lower limits were used in cases of poorly-constrained ages. For systems without estimated ages, we use the simple scaling law presented in \citet{zahnle17}:
\begin{equation}
    I_{XUV}=\frac{I}{I_{\oplus}}\left(\frac{L_{\star}}{L_{\odot}}{}\right)^{-0.6},
\end{equation}
where $I_{XUV}$ is the estimated cumulative XUV flux relative to Earth, $I$ is the planet's present-day instellation (relative to Earth), and $L_{\star}$ is the luminosity of the host star (relative to the Sun).  This approach was consistent within a factor of $\sim2.5$ with our time-evolution model for all planets.

There is large uncertainty in the time evolution of both bolometric and XUV fluxes, especially for the low mass M stars in this work (e.g., \citealt{france22,diamond24}). The values shown in Fig. \ref{fig:shoreline} and \ref{fig:cycle123} should be seen as rough estimations.

\begin{deluxetable*}{lcccccc}
\tablecaption{System Age Values Used in this Study}\label{tab:ages}
\tablewidth{0pt}
\tablehead{\colhead{System} &
\colhead{Age (Gyr)} & \colhead{Reference} }

\startdata
TRAPPIST-1 & $7.6$ &  \citet{burgasser17}\\
GJ 1132 & $6.31$&  \citet{gaidos23}\\
GJ 486 & $3.51$  & \citet{gaidos23} \\
LHS 3844 & $7.8$  & \citet{kane20}\\
GJ 1252 & $6.61$  & \citet{gaidos23}\\
GJ 367 & $7.95$ & \citet{gaidos23} \\
LTT 1445 A & $2$ & \citet{rukdee2024x} \\
GJ 357 & $>5$ & \citet{modirrousta20} \\
GJ 806 & $4$ & \citet{palle23} \\
HD 260655 & 3 & \citet{luque22b} \\
LHS 1140 & $>5$ & \citet{dittmann2017} \\
LP 791-18 & $>0.5$ & \citet{crossfield2019} \\
LTT 3780 & 3.1 & \citet{bonfanti2024} \\
TOI-1468 & $>1$ & \citet{chaturvedi2022} \\
TOI-1685 & 1.3 & \citet{burt2024} \\
TOI-1075 & $>2$ & \citet{essack2023} \\
\enddata

\end{deluxetable*}

\bibliographystyle{aasjournal}
\bibliography{main-v2}

\end{document}